\documentclass[11pt]{article}
\usepackage{graphicx,subfigure,ifthen,color,algorithm,palatino}
\usepackage{amsmath,amsthm,amssymb,amsfonts,verbatim}
\usepackage{mathrsfs,accents,setspace}
\usepackage[autostyle]{csquotes}
\DeclareMathAlphabet\mathbfcal{OMS}{cmsy}{b}{n}
\newboolean{UnBlinded}
\newboolean{DoubleSpaced}
\newboolean{ColourFigures}
\setboolean{UnBlinded}{true}
\setboolean{DoubleSpaced}{false}
\setboolean{ColourFigures}{false}
\definecolor{ruppertgreen}{rgb}{0,.7,.25}

\definecolor{DarkOrange}{rgb}{1,0.549,0}

\newcounter{GAP}

\def\smalldot{\mbox{\fontsize{2mm}{5em}\selectfont{$\bullet$}}}
\newcommand*{\bdotover}[1]{\accentset{\mbox{\smalldot}}{#1}}
\newcommand*{\circover}[1]{\accentset{\circ}{#1}}
\newcommand{\xNonLinearForm}[2]{\bdotover{x}^{#1}_{#2}}
\newcommand{\xLinearForm}[2]{\circover{x}^{#1}_{#2}}
\newcommand{\bxNonLinearForm}[2]{\bdotover{\bx}^{#1}_{#2}}
\newcommand{\bxLinearForm}[2]{\circover{\bx}^{#1}_{#2}}

\newcommand{\bbetaLinearForm}[1]{\circover{\bbeta}^{#1}}
\makeatletter
\def\widebreve{\mathpalette\wide@breve}
\def\wide@breve#1#2{\sbox\z@{$#1#2$}%
     \mathop{\vbox{\m@th\ialign{##\crcr
\kern0.08em\brevefill#1{0.8\wd\z@}\crcr\noalign{\nointerlineskip}%
                    $\hss#1#2\hss$\crcr}}}\limits}
\def\brevefill#1#2{$\m@th\sbox\tw@{$#1($}%
  \hss\resizebox{#2}{\wd\tw@}{\rotatebox[origin=c]{90}{\upshape(}}\hss$}
\makeatletter
\def\bib{\vskip12pt\par\noindent\hangindent=1 true cm\hangafter=1}
\def\MATRIX{\bPsi}
\def\VECTOR{\bpsi}
\def\bbetaBreveTEMP{\bbetabreve_{\mbox{\tiny temp}}}
\def\bd{\boldsymbol{d}}
\def\bx{\boldsymbol{x}}
\def\by{\boldsymbol{y}}
\def\br{\boldsymbol{r}}
\def\bu{\boldsymbol{u}}
\def\bv{\boldsymbol{v}}
\def\bw{\boldsymbol{w}}
\def\Psc{{\mathcal P}}
\def\bg{\boldsymbol{g}}
\def\bdf{\boldsymbol{f}}
\def\bfhat{{\widehat \bdf}}
\def\bpsi{\boldsymbol{\psi}}
\def\real{{\mathbb R}}
\def\balpha{\boldsymbol{\alpha}}
\def\bomega{\boldsymbol{\omega}}
\def\bdeta{\boldsymbol{\eta}}
\def\bdetahat{{\widehat\bdeta}}
\def\expit{\mbox{expit}}
\def\bLambda{\boldsymbol{\Lambda}}
\def\bmu{\boldsymbol{\mu}}
\def\punder{\underline{p}}
\def\bC{\boldsymbol{C}}
\def\bOmega{\boldsymbol{\Omega}}
\def\bUC{\bU_{\mbox{\tiny C}}}
\def\bVC{\bV_{\mbox{\tiny C}}}
\def\bdC{\bd_{\mbox{\tiny C}}}
\def\bsigma{\boldsymbol{\sigma}}
\def\Tsc{{\mathcal T}}

\def\bPsi{\boldsymbol{\Psi}}
\def\bnuubreve{\bpsi_{\ubreve}}
\def\beps{b_{\varepsilon}}
\def\seps{s_{\varepsilon}}
\def\relstack#1#2{\mathop{#1}\limits_{#2}}
\def\MATRIXu{\MATRIX_u}
\def\bru{\br_u}
\def\naturalNumbers{{\mathbb N}}
\def\MATRIXubreve{\bPsi_{\ubreve}}
\def\Nburn{N_{\tiny{\mbox{burn}}}}
\def\Nkept{N_{\tiny{\mbox{kept}}}}
\def\NGibbs{N_{\tiny{\mbox{Gibbs}}}}

\def\bI{\boldsymbol{I}}

\def\bU{\boldsymbol{U}}
\def\bV{\boldsymbol{V}}
\def\br{\boldsymbol{r}}
\def\bz{\boldsymbol{z}}
\def\bxgcurr{\bx_{\bg}^{\mbox{\tiny curr}}}
\def\betabrevecurr{\betabreve ^{\,\mbox{\tiny curr}}}
\def\bUqbetau{\bU_{\qDens(\bbeta,\bu)}}
\def\bdqbetau{\bd_{\qDens(\bbeta,\bu)}}
\def\bSigmaqbetau{\bSigma_{\qDens(\bbeta,\bu)}}
\def\MATRIXbeta{\bPsi_{\beta}}
\def\brbeta{\br_{\beta}}
\def\bnubetabreve{\VECTOR_{\betabreve}}
\def\bnubetabreveFORsqrt{\VECTOR_{\tiny{\betabreve}}}
\def\bdbeta{\bd_{\beta}}
\def\bUbeta{\bU_{\beta}}
\def\bzbeta{\bz_{\beta}}
\def\bzbetabreve{\bz_{\betabreve}}
\def\bzubreve{\bz_{\ubreve}}
\def\bmuqbetau{\bmu_{\qDens(\bbeta,\bu)}}
\def\bmuqu{\bmu_{\qDens(\bu)}}
\def\bSigmaqu{\bSigma_{\qDens(\bu)}}
\def\bCbreve{{\widebreve\bC}}
\def\bCbreveT{{\widebreve\bC}{\null}^T}
\def\bCbrevegT{{\widebreve\bC}{\,\null}^T_{\!\!\bg}}
\def\bUu{\bU_u}
\def\bdu{\bd_u}
\def\bzu{\bz_u}
\def\betabreve{{\widebreve\beta}}
\def\ubreve{{\widebreve u}}
\def\XbreveTalpha{\mbox{${\widebreve {\rm X}}$T$\alpha$}}
\def\ZbreveTalpha{\mbox{${\widebreve {\rm Z}}$T$\alpha$}}
\def\XTy{\textrm{XTy}}
\def\ZTy{\textrm{ZTy}}
\def\XTX{\textrm{XTX}}
\def\ZTX{\textrm{ZTX}}
\def\ZTZ{\textrm{ZTZ}}
\def\CTy{\textrm{CTy}}
\def\CTC{\textrm{CTC}}
\def\bsigsqqbetaubreve{\bsigma^2_{\qDens(\mbox{\tiny{$\bbetabreve,\bubreve$}})}}
\def\bdCsq{\bdC^2}
\def\bmuqbetaubreve{\bmu_{\qDens(\bbetabreve,\bubreve)}}
\def\bmuqubreve{\bmu_{\qDens(\bubreve)}}
\def\bsigsqqubreve{\bsigma^2_{\qDens(\bubreve)}}
\def\bsigsqqbetabreve{\bsigma^2_{\qDens(\bbetabreve)}}
\def\bOmegaqbetau{\bOmega_{\qDens(\bbeta,\bu)}}
\def\XbreveTy{\mbox{${\widebreve {\rm X}}$Ty}}
\def\ZbreveTy{\mbox{${\widebreve {\rm Z}}$Ty}}
\def\CbreveTy{\mbox{${\widebreve {\rm C}}$Ty}}
\def\ZbreveTXbreve{{\mbox{${\widebreve {\rm Z}}$T\mbox{${\widebreve {\rm X}}$}}}}
\def\bXbreveT{{\widebreve\bX}{\null}^T}
\def\bZbreveT{{\widebreve\bZ}{\null}^T}
\def\bubreve{{\widebreve\bu}}
\def\brbetabreve{\br_{\betabreve}}
\def\bomegabetabreve{\bomega_{\betabreve}}
\def\bomegabetabreve{\bomega_{\betabreve}}
\def\brubreve{\br_{\ubreve}}
\def\bomegaubreve{\bomega_{\ubreve}}
\def\GammaDist{\mbox{Gamma}}
\def\bUX{\bU_{\mbox{\tiny{$\!\!\bX$}}}}
\def\bVX{\bV_{\mbox{\tiny{$\!\!\bX$}}}}
\def\bdX{\bd_{\mbox{\tiny{$\bX$}}}}
\def\bUZ{\bU_{\mbox{\tiny{$\!\!\bZ$}}}}
\def\bVZ{\bV_{\mbox{\tiny{$\!\!\bZ$}}}}
\def\bdZ{\bd_{\mbox{\tiny{$\bZ$}}}}
\def\bdXsq{\bdX^2}
\def\bdZsq{\bdZ^2}
\def\bXbreve{{\widebreve\bX}}
\def\bbetabreve{{\widebreve\bbeta}}
\def\bZbreve{{\widebreve\bZ}}
\def\su{s_u}
\def\Cov{\mbox{\rm Cov}}
\def\bX{\boldsymbol{X}}
\def\bZ{\boldsymbol{Z}}
\def\tr{\mbox{tr}}
\def\diag{\mbox{diag}}
\def\bbeta{\boldsymbol{\beta}}
\def\bSigma{\boldsymbol{\Sigma}}
\def\bzero{\boldsymbol{0}}
\def\smhalf{{\textstyle{\frac{1}{2}}}}
\def\simind{\stackrel{{\tiny \mbox{ind.}}}{\sim}}
\def\bone{\boldsymbol{1}}
\def\sigeps{\sigma_{\varepsilon}}
\def\sigsqeps{\sigeps^2}
\def\myand{\&\ }
\def\sigmabeta{\sigma_{\beta}}
\def\pDens{\mathfrak{p}}
\def\qDens{\mathfrak{q}}
\def\epsToler{\varepsilon_{\mbox{\tiny{toler.}}}}
\def\rest{\mbox{rest}}
\def\bxLini{\bxLinearForm{\null}{i}}
\def\bxNoni{\bxNonLinearForm{\null}{i}}
\def\xLinji{\xLinearForm{\null}{ji}}
\def\xNonji{\xNonLinearForm{\null}{ji}}

\def\bxNon{\bxNonLinearForm{\null}{\null}}
\def\bxLiniT{\bxLini{\null^T}}
\def\bxNoniT{\bxNoni{\null^T}}
\def\bxLin{\bxLinearForm{\null}{\null}}
\def\bbetaLin{\bbetaLinearForm{\null}}
\def\bbetaLins{\bbetaLinearForm{[s]}}
\def\dNon{d_{\bullet}}
\def\dLin{d_{\circ}}
\def\bUZj{\bU_{\mbox{\tiny{$\!\!\bZ{j}$}}}}
\def\bVZj{\bV_{\mbox{\tiny{$\!\!\bZ{j}$}}}}
\def\bdZj{\bd_{\mbox{\tiny{$\bZ{j}$}}}}
\def\gothicc{\mathfrak{c}}
\def\XbreveTyadj{\mbox{${\widebreve {\rm X}}$}
                {\rm Ty}_{\mbox{\scriptsize{adj}}}}
\def\ZbreveTyadj{\mbox{${\widebreve {\rm Z}}$}
                {\rm Ty}_{\mbox{\scriptsize{adj}}}}

\def\ZbreveTXbrevej{\textrm{\ZbreveTXbreve}\null
                      ^{\mbox{\scriptsize{$\langle j\rangle$}}}}

\def\ZbreveTZbrevejjd{\textrm{\ZbreveTZbreve}\null
                      ^{\mbox{\scriptsize{$\langle j,j'\rangle$}}}}

\def\ZbreveTXbrevejT{\textrm{\ZbreveTXbreve}{\null}^{\langle j\rangle{T}}}

\def\ZbreveTZbreve{{\mbox{${\widebreve {\rm Z}}$T\mbox{${\widebreve {\rm Z}}$}}}}
\def\ZbreveTXbreve{{\mbox{${\widebreve {\rm Z}}$T\mbox{${\widebreve {\rm X}}$}}}}
\def\bdZjsq{\bdZj^2}

\def\bubrevejCurr{\bubreve_j^{\mbox{\tiny{curr}}}}
\def\brujbreve{\br_{\ubreve{j}}}

\def\bubrevejdCurr{\bubreve_{j'}^{\mbox{\tiny{curr}}}}
\def\bnuujbreve{\bpsi_{\ubreve{j}}}
\def\bzujbreve{\bz_{\ubreve{j}}}
\def\buGbl{\bu_{\mbox{\rm\tiny gbl}}}
\def\buLini{\bu_{\mbox{\rm\tiny lin,$i$}}}
\def\buGrpi{\bu_{\mbox{\rm\tiny grp,$i$}}}
\def\bZgbli{\bZ_{\mbox{\rm\tiny gbl,$i$}}}
\def\bZgrpi{\bZ_{\mbox{\rm\tiny grp,$i$}}}
\def\sigmaGbl{\sigma_{\mbox{\rm\tiny gbl}}}
\def\Kgbl{K_{\mbox{\rm\tiny gbl}}}
\def\sigmaGrp{\sigma_{\mbox{\rm\tiny grp}}}
\def\Kgrp{K_{\mbox{\rm\tiny grp}}}
\def\bGbl{b_{\mbox{\rm\tiny gbl}}}
\def\bGrp{b_{\mbox{\rm\tiny grp}}}
\def\sGbl{s_{\mbox{\rm\tiny gbl}}}
\def\sGrp{s_{\mbox{\rm\tiny grp}}}
\def\sLin{s_{\mbox{\rm\tiny lin}}}
\def\bLino{b_{\mbox{\rm\tiny lin,$1$}}}
\def\bLint{b_{\mbox{\rm\tiny lin,$2$}}}
\def\zgblo{z_{\mbox{\rm\tiny gbl,$1$}}}
\def\zgblk{z_{\mbox{\rm\tiny gbl,$k$}}}
\def\zgblKgbl{z_{\mbox{\rm\tiny gbl,$\Kgbl$}}}

\def\zgrpo{z_{\mbox{\rm\tiny grp,1}}}
\def\zgrpk{z_{\mbox{\rm\tiny grp,k}}}
\def\zgrpKgrp{z_{\mbox{\rm\tiny grp,$\Kgrp$}}}
\def\zgblKgbl{z_{\mbox{\rm\tiny gbl,$\Kgbl$}}}
\def\bZgbl{\bZ_{\mbox{\rm\tiny gbl}}}
\def\bZgblg{\bZ_{\mbox{\rm\tiny gbl},\bg}}
\def\bZgrpg{\bZ_{\mbox{\rm\tiny grp},\bg}}
\def\bZbrevegblg{\bZbreve_{\mbox{\rm\tiny gbl},\bg}}
\def\bZbrevegrpig{\bZbreve_{\mbox{\rm\tiny grp},i,\bg}}
\def\bZgblBreve{\bZbreve_{\text{\tiny gbl}}}
\def\bZgbliBreve{\bZbreve_{\text{\tiny gbl,$i$}}}
\def\bZgbloneBreve{\bZbreve_{\text{\tiny gbl,$1$}}}
\def\bZgblmBreve{\bZbreve_{\text{\tiny gbl,$m$}}}
\def\bUZgbl{\bU_{\tiny{\bZgbl}}}
\def\bUZgrpi{\bU_{\mbox{\tiny{$\bZgrpi$}}}}
\def\bVZgbl{\bV_{\tiny{\bZgbl}}}
\def\bVZgrpi{\bV_{\tiny{\bZgrpi}}}
\def\bdZgbl{\bd_{\tiny{\bZgbl}}}
\def\bdZgblsq{\bdZgbl^2}
\def\XTyiadj{(\mbox{X}{\rm Ty}_{\mbox{\tiny $i$}})_{\mbox{\tiny adj}}}
\def\XTXi{\textrm{XTX}_i}
\def\XTXbrevei{\mbox{XT${\widebreve {\rm X}}$}_i}
\def\XTZgbliBreve{{\mbox{XT\mbox{${\widebreve {\rm Z}}$}}}_{\mbox{\tiny gbl,$i$}}}
\def\XTZgrpiBreve{{\mbox{XT\mbox{${\widebreve {\rm Z}}$}}}_{\mbox{\tiny grp,$i$}}}
\def\XbreveTZgrpiBreve{{\mbox{${\widebreve {\rm X}}$T\mbox{${\widebreve {\rm Z}}$}}}_{\mbox{\tiny grp,$i$}}}

\def\ZgblBreveTZgrpiBreve{\mbox{${\widebreve{\rm Z}}$}_{\mbox{\tiny gbl}}
{\mbox{T${\widebreve{\rm Z}}$}}_{\mbox{\tiny grp,$i$}}}
\def\ZgrpBreveTyiadj{(\mbox{$\widebreve{\rm Z}_{\mbox{\tiny grp}}$}
{\rm Ty}_{\mbox{\tiny $i$}})_{\mbox{\tiny adj}}}
\def\SUM{\rm SUM}

\def\SUMXbreveTyadj{$\SUM$\mbox{${\widebreve {\rm X}}$}
                {\rm Ty}_{\mbox{\scriptsize{adj}}}}
\def\SUMXbreveTZgblBreve{$\SUM${\mbox{${\widebreve {\rm X}}$T
                         \mbox{${\widebreve {\rm Z}}$}}}_{\mbox{\tiny gbl}}}
\def\SUMZgblBreveTyadj{$\SUM$\mbox{$\widebreve{\rm Z}_{\mbox{\tiny gbl}}$}
                      {\rm Ty}_{\mbox{\scriptsize{adj}}}}

\def\bdZgrpi{\bd_{\tiny{\bZgrpi}}}
\def\bdZgrpisq{\bdZgrpi^2}
\def\bZgrpiBreve{\bZbreve_{\text{\tiny grp},i}}
\def\bugblBreve{\bubreve_{\text{\tiny gbl}}}
\def\buGrpiBreve{\bubreve_{\mbox{\rm\tiny grp,$i$}}}
\def\bUuLini{\bU_{\uLini}}
\def\brugblbreve{\br_{\ugblbreve}}
\def\brugrpibreve{\br_{\ugrpibreve}}
\def\bruLini{\br_{\uLini}}

\def\ugblbreve{\ubreve_{\text{\tiny gbl}}}

\def\ugrpibreve{\ubreve_{\text{\tiny grp,$i$}}}

\def\uLini{u_{\text{\tiny lin,$i$}}}
\def\bnuugblbreve{\VECTOR_{\ugblbreve}}

\def\bzugblbreve{\bz_{\ugblbreve}}
\def\bzugrpibreve{\bz_{\ugrpibreve}}
\def\bzuLini{\bz_{\uLini}}

\def\MATRIXuLini{\MATRIX_{\uLini}}
\def\bduLini{\bd_{\uLini}}
\def\bnuugrpibreve{\VECTOR_{\ugrpibreve}}

\def\Ngrid{N_{\tiny{\mbox{grid}}}}

\begin{document}

\ifthenelse{\boolean{DoubleSpaced}}{\doublespacing}{}

\vskip5mm
\centerline{\Large\bf Orthogonalized Design Matrices Speed-ups of}      

\vskip1mm
\centerline{\Large\bf Bayesian Semiparametric Regression}  
      
\vskip5mm
\centerline{\normalsize\sc Nurul Fitriyani and Matt P. Wand}
\vskip5mm
\centerline{\textit{School of Mathematical and Physical Sciences,
University of Technology Sydney}}
\vskip5mm
\centerline{9th July, 2026}
\vskip5mm
\centerline{\large\bf Abstract}
\vskip2mm

We explain how important classes of Bayesian semiparametric
regression fitting and inference procedures
can be sped up, significantly, via the use of orthogonalized design
matrices. Typically, design matrices in semiparametric regression contain
predictor observations and basis functions of such data. 
In \emph{Bayesian} semiparametric regression,
loop-type approaches such as Gibbs sampling and 
coordinate ascent variational inference typically are required. 
We show that pre-loop reformulation of Bayesian semiparametric
regression models involving orthogonalized design matrices 
lead to two orders of magnitude, with respect to 
column dimension, computational reduction. Our computer experiments
reveal that this simple paradigm results in approximately
5- to 60-fold speed-ups.


\section{Introduction}\label{sec:intro}

Practical fitting and inference for Bayesian semiparametric 
regression models usually requires the running of loop-type 
procedures such as Gibbs sampling or coordinate ascent
variational inference. The essence of this article
is explaining and demonstrating how conversion
of the model to an equivalent form involving 
orthogonalized design matrices provides two orders
of magnitude reductions in computation.
These orders of magnitude are with respect to the
column dimensions of the design matrices, which
can be in the dozens or even hundreds for 
contemporary Bayesian semiparametric models.
Computer experiments reveal that practical speed-ups 
are in the approximate 5--60 factor range. 

In Section \ref{sec:bayNonParReg} we commence our explanation
of orthogonalized design matrices speed-ups for a Bayesian
nonparametric regression model. This allows us to describe
the essence of the approach with minimal notational overhead.
Both Gaussian and Bernoulli response models are shown
to benefit from the approach and our computer experiments
in this section demonstrate speed-ups as high as a factor
of around 60. We also corroborate these results with
some order of magnitude comparisons. Algorithms \ref{alg:odmGibbsPenSpl}
and \ref{alg:odmProbitGibbsPenSpl} facilitate implementation.

Sections \ref{sec:GAM} and \ref{sec:GSCmodels} explore the central theme applied
to generalized additive models and group-specific curves models, respectively.
For such arbitrarily large models the speed-ups offered by use
orthogonalized design matrices can have noticeable practical benefits,
as demonstrated by some simulated and actual data examples.

Even though speeding up Gibbs sampling is the main focus of this
article, the general approach also applies to other Bayesian
inference approach such as coordinate ascent variational inference.
In Section \ref{sec:varInf} we provide illustration for the
Bayesian nonparametric regression model from Section \ref{sec:bayNonParReg}.

In Sections \ref{sec:bayNonParReg}--\ref{sec:varInf} we present
five algorithms with orthogonalized design matrices speed-ups
across three Bayesian semiparametric regression model types.
These allow concrete illustration and evaluation
of the approach. However, orthogonalized design matrices speed-ups 
is a general paradigm that applies to many other Bayesian 
regression-type models.

Section \ref{sec:conclusion} contains some concluding discussion.

\subsection{Notation}

Scalar functions applied to a vector are evaluated
in an element-wise fashion. For example, 
$\cosh([3\ 11]^T)\equiv[\cosh(3)\ \cosh(11)]^T$.
If $\bv$ is a column vector then
$\Vert\bv\Vert\equiv\sqrt{\bv^T\bv}$ is
the Euclidean norm of $\bv$. 
The notation $\diag(\bv)$ is used for the diagonal
matrix containing the entries of $\bv$ along its
diagonal. If $\bv$ and $\bw$ are column vectors
of the same size then $\bv\odot\bw$ and $\bv/\bw$ 
are the column vectors of element-wise products
and quotients, respectively. Also, $\bone$ is a 
column vector of ones. The symbol $\simind$ is shorthand 
for ``independently distributed as''. The random variable
$x$ has a Gamma distribution with shape parameter $\kappa>0$
and rate parameter $\lambda>0$, written $x\sim\mbox{Gamma}(\kappa,\lambda)$,
if and only if the density function of $x$ is 
$\pDens(x)=\{\lambda^{\kappa}/\Gamma(\kappa)\}\,x^{\kappa-1}\exp(-\lambda\,x),
\ x>0.$ 
The random variable
$x$ has an Inverse Gamma distribution with shape parameter $\kappa>0$
and rate parameter $\lambda>0$, written $x\sim\mbox{Inverse-Gamma}(\kappa,\lambda)$,
if and only if the density function of $x$ is 
$\pDens(x)=\{\lambda^{\kappa}/\Gamma(\kappa)\}\,x^{-\kappa-1}\exp(-\lambda/x)$,
$x>0.$  The symbol $\Phi$ is used for cumulative distribution function of 
the $N(0,1)$ distribution.
For a logical proposition $\Psc$, $I(\Psc)=1$ if
$\Psc$ is true and $I(\Psc)=0$ if $\Psc$ does not hold.

\section{Nonparametric Regression}\label{sec:bayNonParReg}

Bayesian nonparametric regression via penalized splines
(e.g.\ Harezlak \textit{et al.}, 2018, Chapter 2)
is one of the simplest model types
that benefits from the use of orthogonalized design matrices.
In this section we start with the simplest Gaussian
response case and then, later, discuss other response situations.

\subsection{Gaussian Responses}

For univariate and continuous predictor/response
pairs $(x_i,y_i)$, $1\le i\le n$, the Gaussian responses 
nonparametric regression model has the generic form
\begin{equation}
y_i\simind N(f(x_i),\sigeps^2),\quad 1\le i\le n.
\label{eq:GaussNPReg}
\end{equation}
Throughout this section we consider the following
Bayesian penalized spline model for $f$:
$$f(x)=\beta_0+\beta_1\,x+\sum_{k=1}^K u_k z_k(x),
\quad u_k|\sigma_u^2\simind N(0,\sigma_u^2),
$$
where $\{z_k(\cdot):1\le k\le K\}$ is a suitable
spline basis. Typically $K$ is an integer between
around $25$ and $50$, but may be higher if $f$
is thought to be particularly wiggly
(e.g.\ Section 2.4 of Harezlak \textit{et al.}, 2018).
A recommended default choice of the $z_k$ 
is described in Section 4 of Wand \myand Ormerod (2008)
and corresponds to low-rank smoothing splines.
The full description of the model that we consider
here is
\begin{equation}
\begin{array}{c}
\by|\bbeta, \bu, \sigsqeps \sim N(\bX\bbeta+\bZ\,\bu,\sigeps^2\,\bI_n),
\quad
\bbeta\sim N(\bzero,\sigmabeta^2\bI_2),
\quad \bu|\sigma_u^2
\sim N(\bzero,\sigma_u^2\bI_K),\\[1ex]
\quad\sigma_u^{-2}|b_u\sim\mbox{Gamma}\big(\smhalf,b_u\big),
\quad b_u\sim\mbox{Gamma}(\smhalf,s_u^{-2}),\\[1ex]
\quad\sigeps^{-2}|\beps\sim\mbox{Gamma}(\smhalf,\beps),
\quad\beps\sim\mbox{Gamma}(\smhalf,\seps^{-2})
\end{array}
\label{eq:BayePenSplineModel}
\end{equation}
where $\sigma_{\beta},\su,\seps>0$ are user-specified hyperparameters,
$$\by\equiv[y_i]_{1\le i\le n},\quad
\bX\equiv[1\ x_i]_{1\le i\le n}\quad\mbox{and}\quad
\bZ\equiv[\relstack{z_k(x_i)}{1\le k\le K}]_{1\le i\le n}.
$$
The vectors $\bbeta\ (2\times1)$ and $\bu\ (K\times1)$ contain
the $\beta_k$ and $u_k$, respectively. The distributional structure
of $\sigeps$ in (\ref{eq:BayePenSplineModel}) involving the
auxiliary variable $\beps$ is equivalent to imposition of the
prior density function
$$\pDens(\sigeps)=\frac{2}{\pi\{1+(\sigeps^2/\seps^2)\}\seps},
\quad\sigeps>0,$$
which corresponds to the Half-Cauchy distribution with scale
parameter $\seps$ (e.g. Gelman, 2006). The prior on 
$\sigma_u$ is analogous.

Standard calculations show that the full conditional
distribution of $\bu$ is 
{\setlength\arraycolsep{1pt}
\begin{eqnarray*}
&&\bu|\rest\sim N\big(\MATRIXu^{-1}\sigeps^{-2}\bru,\MATRIXu^{-1})
\quad \mbox{where}\quad\bru\equiv\bZ^T(\by-\bX\bbeta)\\[0ex]
&&\qquad\qquad\mbox{and}\quad
\MATRIXu\equiv \sigeps^{-2}\bZ^T\bZ+\sigma_u^{-2}\bI_K
\end{eqnarray*}
}
where `rest' denotes all random variables in (\ref{eq:BayePenSplineModel})
other than $\bu$. An analogous result holds for $\bbeta|\rest$.
Derivations of the full conditionals of the scalar variables
in (\ref{eq:BayePenSplineModel}) are particular simple, and
lead to the direct Gibbs sampling scheme listed in 
Algorithm \ref{alg:dirGibbsPenSpl}.
Result 1 in the appendix justifies the forms of the
$\bbeta$ and $\bu$ Multivariate Normal draws.
The main output of Algorithm \ref{alg:dirGibbsPenSpl} is
the kept Gibbs samples of fit vectors
for inputted grid-wise design matrices $\bX_{\bg}$
and $\bZ_{\bg}$. These matrices are defined analogously
to $\bX$ and $\bZ$ but with basis functions evaluated
at grid points stored in an arbitrary vector $\Ngrid$ of plotting 
abscissae, $\bg$, rather than the $x_i$s.

\begin{algorithm}[t]
\begin{center}
\begin{minipage}[t]{155mm}
\hrule
\begin{small}
\begin{itemize}
\setlength\itemsep{0pt}
\item[] Inputs: $\by$ $(n\times 1)$,\ $\bX$ $(n\times2)$,\ $\bZ$ $(n\times K)$,
\ $\bX_{\bg}$ $(\Ngrid\times2)$,\ $\bZ_{\bg}$ $(\Ngrid\times K)$,
\item[] $\qquad\quad\ \ $$\sigmabeta,\su,\seps>0$,\ $\Nburn,\Nkept\in\naturalNumbers$.
\item[] Initialize: $\bu^{[0]}$ $(K\times 1)$, $(\sigeps^{-2})^{[0]},\beps,
(\sigma_u^{-2})^{[0]},b_u>0$.
\item[] 
$\XTy\longleftarrow \bX^T\by$\ ;\ $\ZTy\longleftarrow \bZ^T\by$
\ ;\ $\XTX\longleftarrow \bX^T\bX$
\ ;\ $\ZTX\longleftarrow \bZ^T\bX$
\ ;\ $\ZTZ\longleftarrow \bZ^T\bZ$
\item[] For $s=1,\ldots,\Nburn+\Nkept$:
\begin{itemize}
\item[] $\brbeta\longleftarrow\XTy-\ZTX^T\,\bu^{[s-1]}$\ \ ;\ \ 
$\MATRIXbeta\longleftarrow (\sigeps^{-2})^{[s-1]}\XTX+\sigmabeta^{-2}\bI_2$
\item[] Decompose $\MATRIXbeta=\bUbeta\diag(\bdbeta)\bUbeta^T$ where 
$\bUbeta^T\bUbeta=\bUbeta\bUbeta^T=\bI_2$
\item[] $\bzbeta\sim N(\bzero,\bI_2)$\ \ \ ;\ \ \ 
$\bbeta^{[s]}\longleftarrow\bUbeta\left({\displaystyle\frac{
\bUbeta^T\bzbeta}
{\sqrt{\bdbeta}}}+{\displaystyle\frac{
(\sigeps^{-2})^{[s-1]}
\bUbeta^T\brbeta}
{\bdbeta}}\right)$
\item[] $\bru\longleftarrow \ZTy-\ZTX\,\bbeta^{[s]}$\ \ ;\ \ 
$\MATRIXu\longleftarrow (\sigeps^{-2})^{[s-1]}\ZTZ+(\sigma_u^{-2})^{[s-1]}\bI_K$
\item[] Decompose $\MATRIXu=\bUu\diag(\bdu)\bUu^T$ where 
$\bUu^T\bUu=\bUu\bUu^T=\bI_K$ 
\item[] $\bzu\sim N(\bzero,\bI_K)$\ \ \ ;\ \ \ 
$\bu^{[s]}\longleftarrow\bUu\left({\displaystyle\frac{\bUu^T\bzu}
{\sqrt{\bdu}}}+{\displaystyle\frac{(\sigeps^{-2})^{[s-1]}\bUu^T\bru}
{\bdu}}\right)$
\item[] $(\sigma_u^{-2})^{[s]}\longleftarrow\GammaDist\Big(\smhalf(K+1),b_u+
\smhalf\Vert\bu^{[s]}\Vert^2\Big)$
\ ;\ $b_u\longleftarrow\GammaDist\Big(1,(\sigma_u^{-2})^{[s]}+
s_u^{-2}\Big)$
\item[] $(\sigeps^{-2})^{[s]}\longleftarrow\GammaDist\Big(\smhalf(n+1),\beps+
\smhalf\Vert\by-\bX\bbeta^{[s]}-\bZ\bu^{[s]}\Vert^2\Big)$
\item[] $\beps\longleftarrow\GammaDist\Big(1,(\sigeps^{-2})^{[s]}+
\seps^{-2}\Big)$
\end{itemize}
\item[] For $s=1,\ldots,\Nkept$:
\begin{itemize}
\item[] $\bfhat_{\bg}^{[s]}\longleftarrow\bX_{\bg}\bbeta^{[s+\Nburn]}
+\bZ_{\bg}\bu^{[s+\Nburn]}$
\item[] $(\sigeps^{-2})^{[s]}\longleftarrow (\sigeps^{-2})^{[s+\Nburn]}$
\ \ ;\ \ 
$(\sigma_u^{-2})^{[s]}\longleftarrow (\sigma_u^{-2})^{[s+\Nburn]}$
\end{itemize}
\item[] Outputs: $\big\{\bfhat_{\bg}^{[s]},(\sigeps^{-2})^{[s]},
(\sigma_u^{-2})^{[s]}:1\le s\le \Nkept\big\}$
\end{itemize}
\end{small}
\hrule
\end{minipage}
\end{center}
\caption{\textit{A direct Gibbs sampling algorithm 
for the Bayesian nonparametric regression model 
(\ref{eq:BayePenSplineModel}).}}
\label{alg:dirGibbsPenSpl}
\end{algorithm}

\begin{algorithm}[t]
\begin{center}
\begin{minipage}[t]{155mm}
\hrule
\begin{small}
\begin{itemize}
\setlength\itemsep{0pt}
\item[] Inputs: $\by$ $(n\times 1)$,\ $\bX$ $(n\times2)$,\ $\bZ$ $(n\times K)$,
\ $\bX_{\bg}$ $(\Ngrid\times2)$,\ $\bZ_{\bg}$ $(\Ngrid\times K)$,
\item[] $\qquad\quad\ \ $$\sigmabeta,\su,\seps>0$,\ $\Nburn,\Nkept\in\naturalNumbers$.
\item[] Decompose $\bX=\bUX\diag\big(\bdX\big)\bVX^T$ where 
$\bUX^T\bUX=\bVX^T\bVX=\bVX\bVX^T=\bI_2$
\item[] Decompose $\bZ=\bUZ\diag\big(\bdZ\big)\bVZ^T$ where 
$\bUZ^T\bUZ=\bVZ^T\bVZ=\bVZ\bVZ^T=\bI_K$
\item[] $\bXbreve\longleftarrow\bUX\diag(\bdX)$\ \ ;\ \ 
$\bZbreve\longleftarrow\bUZ\diag(\bdZ)$
\ \ ;\ \ $\bdXsq\longleftarrow\bdX\odot\bdX$
\ \ ;\ \ $\bdZsq\longleftarrow\bdZ\odot\bdZ$
\item[] $\XbreveTy\longleftarrow \bXbreveT\by$
\ ;\ $\ZbreveTy\longleftarrow \bZbreveT\by$
\ ;\ $\ZbreveTXbreve\longleftarrow\bZbreveT\bXbreve$
\item[] Initialize: $\bubreve^{[0]}$ $(K\times 1)$, 
$(\sigeps^{-2})^{[0]},\beps,
(\sigma_u^{-2})^{[0]},b_u>0$
\item[] For $s=1,\ldots,\Nburn+\Nkept$: 
\begin{itemize}
\item[] $\brbetabreve\longleftarrow 
\XbreveTy-(\ZbreveTXbreve)^T\,\bubreve^{[s-1]}$\ \ ;\ \ 
$\bnubetabreve\longleftarrow(\sigeps^{-2})^{[s-1]}
\bdX^2+\sigmabeta^{-2}\bone_2$
\item[] $\bzbetabreve\sim N(\bzero,\bI_2)$\ \ \ ;\ \ \ 
$\bbetabreve^{[s]}\longleftarrow {\displaystyle\frac{\bzbetabreve}
{\sqrt{\bnubetabreveFORsqrt}}}+{\displaystyle\frac{(\sigeps^{-2})^{[s-1]}
\brbetabreve}
{\bnubetabreve}}$
\item[] $\brubreve\longleftarrow 
\ZbreveTy-\ZbreveTXbreve\,\bbetabreve^{[s]}$\ \ ;\ \ 
$\bnuubreve\longleftarrow(\sigeps^{-2})^{[s-1]}
\bdZ^2+(\sigma_u^{-2})^{[s-1]}\bone_K$
\item[] $\bzubreve\sim N(\bzero,\bI_K)$\ \ \ ;\ \ \ 
$\bubreve^{[s]}\longleftarrow {\displaystyle\frac{\bzubreve}
{\sqrt{\bnuubreve}}}+{\displaystyle\frac{
(\sigeps^{-2})^{[s-1]}
\brubreve}{\bnuubreve}}$
\item[] $(\sigma_u^{-2})^{[s]}\longleftarrow\GammaDist\Big(\smhalf(K+1),b_u+
\smhalf\big\Vert\bubreve^{[s]}\big\Vert^2\Big)$
\ \ ;\ \ $b_u\longleftarrow\GammaDist\Big(1,(\sigma_u^{-2})^{[s]}+
s_u^{-2}\Big)$
\item[] $(\sigeps^{-2})^{[s]}\longleftarrow\GammaDist\Big(\smhalf(n+1),\beps+
\smhalf\Big\Vert\by-\bXbreve\bbetabreve^{[s]}-\bZbreve\bubreve^{[s]}
\Big\Vert^2\Big)$
\item[] $\beps\longleftarrow\GammaDist\Big(1,(\sigeps^{-2})^{[s]}+
\seps^{-2}\Big)$
\end{itemize}
\item[] $\bXbreve_{\bg}\longleftarrow \bX_{\bg}\bVX$
\ \ \ ;\ \ \ $\bZbreve_{\bg}\longleftarrow \bZ_{\bg}\bVZ$
\item[] For $s=1,\ldots,\Nkept$:
\begin{itemize}
\item[] $\bfhat_{\bg}^{[s]}\longleftarrow\bXbreve_{\bg}\bbetabreve^{[s+\Nburn]}
+\bZbreve_{\bg}\bubreve^{[s+\Nburn]}$
\item[] $(\sigeps^{-2})^{[s]}\longleftarrow (\sigeps^{-2})^{[s+\Nburn]}$
\ \ ;\ \ 
$(\sigma_u^{-2})^{[s]}\longleftarrow (\sigma_u^{-2})^{[s+\Nburn]}$
\end{itemize}
\item[] Outputs: $\big\{\bfhat_{\bg}^{[s]},(\sigeps^{-2})^{[s]},
(\sigma_u^{-2})^{[s]}:1\le s\le \Nkept\big\}$
\end{itemize}
\end{small}
\hrule
\end{minipage}
\end{center}
\caption{\textit{An orthogonalized design matrices speed-up of
Algorithm \ref{alg:dirGibbsPenSpl} for the Bayesian nonparametric
regression model (\ref{eq:BayePenSplineModel}).}}
\label{alg:odmGibbsPenSpl}
\end{algorithm}
%

The main bottleneck in Algorithm \ref{alg:dirGibbsPenSpl} is 
obtaining the singular value decomposition of the $K\times K$ matrix
$\MATRIX_u$ for each iteration of the Gibbs sampling scheme,
which requires $O(K^3)$ operations.
The essence of the approach advocated in this article 
starts with the singular value decomposition of $\bZ$:
\begin{equation}
\bZ=\bUZ\diag\big(\bdZ\big)\bVZ^T
\quad\mbox{where}\quad\bUZ^T\bUZ=\bVZ^T\bVZ=\bVZ\bVZ^T=\bI_K
\label{eq:ghostTrain}
\end{equation}
and where $\bUZ$ is $n\times K$, $\bdZ$ is $K\times1$ and
$\bVZ$ is $K\times K$. Then observe that
\begin{equation}
\bZ\bu=\bZbreve\bubreve
\quad\mbox{where}\quad\bZbreve\equiv\bUZ\diag\big(\bdZ\big)
\quad\mbox{and}\quad\bubreve\equiv \bVZ^T\bu.
\label{eq:riverCaves}
\end{equation}
Noting that
$$\Cov(\bubreve|\sigma_u^2)=\Cov(\bVZ^T\bu|\sigma_u^2)
=\bVZ^T\sigma_u^2\bI_K\,\bVZ=\sigma_u^2\bI_K$$
and applying the same logic to $\bX$ and $\bbeta$,
model (\ref{eq:BayePenSplineModel}) is equivalent
to an alternative formulation for which the
first line is replaced by
$$
\by|\bbetabreve,\bubreve,\sigsqeps 
\sim N(\bXbreve\bbetabreve+\bZbreve\,\bubreve,\sigeps^2\,\bI_n),
\quad
\bbetabreve\sim N(\bzero,\sigmabeta^2\bI_2),
\quad \bubreve|\sigma_u^2
\sim N(\bzero,\sigma_u^2\bI_K).
$$
However, $\bZbreve$ is such that
\begin{equation}
\bZbreveT\bZbreve=\big\{\bUZ\diag\big(\bdZ\big)\big\}^T
\bUZ\diag\big(\bdZ\big)
=\diag\big(\bdZ\big)\bUZ^T\bUZ \diag\big(\bdZ\big)
=\diag\big(\bdZ\odot\bdZ\big)
\label{eq:ZbreveTZbreve}
\end{equation}
which implies that the columns of $\bZbreve$ are orthogonal
vectors in $\real^n$. Use of $\bZbreve$ is loosely related
to the so-called \emph{Demmler-Reinsch basis} version of
smoothing splines (Demmler \myand Reinsch, 1975),
but is simply a linear transformation of $\bZ$ based on 
(\ref{eq:ghostTrain}) and (\ref{eq:riverCaves}).
The full conditional distribution of $\bubreve$ is 
{\setlength\arraycolsep{1pt}
\begin{eqnarray*}
&&\bubreve|\rest\sim N\big(\MATRIXubreve^{-1}\sigeps^{-2}\brubreve,
\MATRIXubreve^{-1})
\quad \mbox{where}\quad\brubreve
\equiv\bZbreveT(\by-\bXbreve\bbetabreve)\\[0.5ex]
&&\qquad\qquad\mbox{and}\quad
\MATRIXubreve\equiv\diag\Big(\sigeps^{-2}(\bdZ\odot\bdZ)
+\sigma_u^{-2}\bone_K\Big)
\end{eqnarray*}
}
The fact that $\MATRIXubreve$ is a diagonal matrix implies that
the need for $K\times K$ matrix inversion is avoided when
obtaining draws from the full conditional distribution of $\bubreve$.
In view of Result 1, the following $O(K)$ steps lead to a draw from
$\bubreve|\rest$:
$$\bnuubreve\longleftarrow \sigeps^{-2}(\bdZ\odot\bdZ)
+\sigma_u^{-2}\bone_K\ \ ;\ \ \bz\sim N(\bzero,\bI_K)\ \ ;\ \ 
\bubreve\longleftarrow\frac{\bz}{\sqrt{\bnuubreve}}+
\frac{\sigeps^{-2}\brubreve}{\bnuubreve}.
$$
Similar statements apply to the $\bbetabreve$ full conditional
draws. Since the construction of $\bZbreve$ is done outside
of the Gibbs sampling loop there is, as a function of $K$, 
a two orders of magnitude reduction in computation realized
by working with the orthogonalized design matrices $\bXbreve$
and $\bZbreve$.

Algorithm \ref{alg:odmGibbsPenSpl} describes this
faster orthogonalized design matrices alternative to 
Algorithm \ref{alg:dirGibbsPenSpl}.

\subsection{Bernoulli Responses with Probit Link}

Now suppose that the $y_i$ are binary: $y_i\in\{0,1\}$.
Then an appropriate alternative to (\ref{eq:GaussNPReg})
is the probit link nonparametric regression model
\begin{equation}
y_i\simind\mbox{Bernoulli}\big(\Phi\big(f(x_i)\big)\big),
\quad 1\le i\le n,
\label{eq:probitNPreg}
\end{equation}
Following Albert \myand Chib (1993), a useful
adaptation of (\ref{eq:BayePenSplineModel}) for fitting 
(\ref{eq:probitNPreg}), involving the auxiliary variables vector $\balpha$,
is
\begin{equation}
\begin{array}{c}
\balpha\ (n\times 1)\ \mbox{has $i$th entry $\alpha_i$
such that $y_i=I\big(\alpha_i\ge0\big)$,\quad $1\le i\le n$,}\\[1ex]
\balpha|\bbeta, \bu \sim N(\bX\bbeta+\bZ\,\bu,\bI_n),
\quad
\bbeta\sim N(\bzero,\sigmabeta^2\bI_2),
\quad \bu|\sigma_u^2
\sim N(\bzero,\sigma_u^2\bI_K),
\\[1ex]
\quad\sigma_u^{-2}|b_u\sim\mbox{Gamma}\big(\smhalf,b_u\big),
\quad b_u\sim\mbox{Gamma}(\smhalf,\su^{-2}),
\end{array}
\label{eq:ProbitPenSplineModel}
\end{equation}
Gibbs sampling for (\ref{eq:ProbitPenSplineModel}) proceeds 
similarly to that laid out in Algorithms \ref{alg:dirGibbsPenSpl}
and \ref{alg:odmGibbsPenSpl}. The main addition is due to
the distributional result
$$(2y_i-1)\,\alpha_i\big|\rest\sim
\mbox{Truncated-Normal}_+\Big((2y_i-1)(\bX\bbeta+\bZ\bu)_i,1\Big),\quad
1\le i\le n,
$$
where the random variable $x$ has a $\mbox{Truncated-Normal}_+$
distribution with location parameter $\mu$ and scale parameter $\sigma$,
written 
$x\sim \mbox{Truncated-Normal}_+(\mu,\sigma^2)$,
if and only if the density function of $x$ is
$$\pDens(x)=\frac{\exp\{-(x-\mu)^2/(2\sigma^2)\}I(x>0)}{
\sigma\Phi(\mu/\sigma)\sqrt{2\pi}}.
$$
Robert (1995) describes methodology for the efficient generation
of $\mbox{Truncated-Normal}_+$ random variables. 
Algorithm \ref{alg:odmProbitGibbsPenSpl} is a resultant
orthogonalized design matrices Gibbs sampling algorithm.

\begin{algorithm}[t]
\begin{center}
\begin{minipage}[t]{155mm}
\hrule
\begin{small}
\begin{itemize}
\setlength\itemsep{0pt}
\item[] Inputs: $\by$ $(n\times 1)$,\ $\bX$ $(n\times2)$,\ $\bZ$ $(n\times K)$, 
\ $\bX_{\bg}$ $(\Ngrid\times2)$,\ $\bZ_{\bg}$ $(\Ngrid\times K)$
\item[] $\qquad\quad\ \ $$\sigmabeta,\su,\seps>0$,\ $\Nburn,\Nkept\in\naturalNumbers$.
\item[] Decompose $\bX=\bUX\diag\big(\bdX\big)\bVX^T$ where 
$\bUX^T\bUX=\bVX^T\bVX=\bVX\bVX^T=\bI_2$
\item[] Decompose $\bZ=\bUZ\diag\big(\bdZ\big)\bVZ^T$ where 
$\bUZ^T\bUZ=\bVZ^T\bVZ=\bVZ\bVZ^T=\bI_K$
\item[] $\bXbreve\longleftarrow\bUX\diag\big(\bdX\big)$\ \ ;\ \ 
$\bZbreve\longleftarrow\bUZ\diag\big(\bdZ\big)$
\ \ ;\ \ $\bdXsq\longleftarrow\bdX\odot\bdX$
\ \ ;\ \ $\bdZsq\longleftarrow\bdZ\odot\bdZ$
\item[] $\ZbreveTXbreve\longleftarrow\bZbreveT\bXbreve$
\item[] Initialize: $\bubreve^{[0]}$ $(K\times 1)$, $(\sigma_u^{-2})^{[0]},b_u>0$,
\ $\XbreveTalpha\ (2\times 1)$,\ $\ZbreveTalpha\ (K\times 1)$
\item[] For $s=1,\ldots,\Nburn+\Nkept$:
\begin{itemize}
\item[] $\bomegabetabreve\longleftarrow 
\XbreveTalpha-(\ZbreveTXbreve)^T\,\bubreve^{[s-1]}$\ \ ;\ \ 
$\bnubetabreve\longleftarrow\bdX^2+\sigmabeta^{-2}\bone_2$
\item[] $\bzbetabreve\sim N(\bzero,\bI_2)$\ \ \ ;\ \ \ 
$\bbetabreve^{[s]}\longleftarrow {\displaystyle\frac{\bzbetabreve}
{\sqrt{\bnubetabreveFORsqrt}}}+{\displaystyle\frac{\bomegabetabreve}
{\bnubetabreve}}$
\item[] $\bomegaubreve\longleftarrow 
\ZbreveTalpha-\ZbreveTXbreve\,\bbetabreve^{[s]}$\ \ ;\ \ 
$\bnuubreve\longleftarrow\bdZ^2+(\sigma_u^{-2})^{[s-1]}\bone_K$
\item[] $\bzubreve\sim N(\bzero,\bI_K)$\ \ \ ;\ \ \ 
$\bubreve^{[s]}\longleftarrow {\displaystyle\frac{\bzubreve}
{\sqrt{\bnuubreve}}}+{\displaystyle\frac{\bomegaubreve}
{\bnuubreve}}$
\item[] $(\sigma_u^{-2})^{[s]}\longleftarrow\GammaDist\Big(\smhalf(K+1),b_u+
\smhalf\big\Vert\bubreve^{[s]}\big\Vert^2\Big)$
\ \ ;\ \ $b_u\longleftarrow\GammaDist\Big(1,(\sigma_u^{-2})^{[s]}+
\su^{-2}\Big)$
\item[] For $i=1,\ldots,n$:
\begin{itemize}
\item[] $\zeta\sim\mbox{Truncated-Normal}_+\big((2y_i-1)
(\bXbreve\bbetabreve^{[s]}+\bZbreve\bubreve^{[s]})_i,1\big)$
\ \ ;\ \ $\alpha_i\longleftarrow (2y_i-1)\zeta$
\end{itemize}
\item[] $\XbreveTalpha\longleftarrow\bXbreveT\balpha$
\ \ ;\ \ $\ZbreveTalpha\longleftarrow\bZbreveT\balpha$
\end{itemize}
\item[] $\bXbreve_{\bg}\longleftarrow \bX_{\bg}\bVX$
\ \ \ ;\ \ \ $\bZbreve_{\bg}\longleftarrow \bZ_{\bg}\bVZ$
\item[] For $s=1,\ldots,\Nkept$:
\begin{itemize}
\item[] $\bdetahat_{\bg}^{[s]}\longleftarrow\bXbreve_{\bg}\bbetabreve^{[s+\Nburn]}
+\bZbreve_{\bg}\bubreve^{[s+\Nburn]}$\ \ \ ;\ \ \  
$(\sigma_u^{-2})^{[s]}\longleftarrow (\sigma_u^{-2})^{[s+\Nburn]}$
\end{itemize}
\item[] Outputs: $\big\{\bdetahat_{\bg}^{[s]},
(\sigma_u^{-2})^{[s]}:1\le s\le \Nkept\big\}$
\end{itemize}
\end{small}
\hrule
\end{minipage}
\end{center}
\caption{\textit{An orthogonalized design matrices Gibbs
sampling scheme for the Bayesian probit nonparametric regression 
model (\ref{eq:ProbitPenSplineModel}).}}
\label{alg:odmProbitGibbsPenSpl}
\end{algorithm}
%

\subsection{Other Links and Response Types}

For binary response Bayesian regression models 
with non-probit links and other response types,
such as counts, the full conditional distributions
involve weighted forms, which nullify the  
advantage of orthogonalized design matrices.
Consider, for example, the Bayesian penalized
spline model with logit link and P\'olya-Gamma
augmentation:
\begin{equation}
\begin{array}{c}
y_i|\bbeta, \bu \simind \mbox{Bernoulli}\big(\expit\{(\bX\bbeta+\bZ\,\bu)_i\}\big),
\quad
\bbeta\sim N(\bzero,\sigmabeta^2\bI_2),
\quad \bu|\sigma_u^2
\sim N(\bzero,\sigma_u^2\bI_K),
\\[1ex]
\quad\sigma_u^{-2}|b_u\sim\mbox{Gamma}\big(\smhalf,b_u\big),
\quad b_u\sim\mbox{Gamma}(\smhalf,\su^{-2}),\\[1ex]
\alpha_i|\bbeta, \bu \simind \mbox{P\'olya-Gamma}(1, (\bX\bbeta+\bZ\,\bu)_i),
\quad 1\le i\le n.
\end{array}
\label{eq:LogisticPenSplineModel}
\end{equation}
Here $\expit(x)\equiv 1/(1+e^{-x})$ and the P\'olya-Gamma
distribution is as defined in Polson \textit{et al.} (2013).
For model (\ref{eq:LogisticPenSplineModel}), the full conditional 
distribution of $\bu$ is
{\setlength\arraycolsep{1pt}
\begin{eqnarray*}
&&\bu|\rest\sim N\big(\MATRIXu^{-1}\bru,\MATRIXu^{-1})
\quad \mbox{where}\quad\bru\equiv\bZ^T
\big\{\by-\smhalf\bone-\balpha\odot(\bX\bbeta)\big\},\\[0ex]
&&\qquad\qquad
\MATRIXu\equiv\bZ^T\diag(\balpha)\bZ+\sigma_u^{-2}\bI_K
\end{eqnarray*}
}
and $\balpha$ is the vector of $\alpha_i$ values.
The generalization of (\ref{eq:ZbreveTZbreve}) to this
weighted case is
$$\bZbreveT\diag(\balpha)\bZbreve=\big\{\bUZ\diag\big(\bdZ\big)\big\}^T
\diag(\balpha)
\bUZ\diag\big(\bdZ\big)
=\diag\big(\bdZ\big)\bUZ^T\diag(\balpha)\bUZ \diag\big(\bdZ\big)
$$
which is not necessarily diagonal. Also, since the $\balpha$
vector changes throughout the Gibbs sampling iterations
there is no re-definition of $\bZbreve$ that leads to
the precision matrix of $\bu|\rest$ having a fixed
diagonal form as a function of $\sigeps^2$ and $\sigma_u^2$.
Similar comments apply to non-Gibbsian approaches such
as those involving Metropolis-Hastings schemes.
Nevertheless, the Gaussian and Bernoulli response
cases are ubiquitous in semiparametric regression
and significantly speeding up their Bayesian analyses 
is worthwhile.

\subsection{Operations Order of Magnitude Comparisons}

Let $\NGibbs\equiv\Nburn+\Nkept$ and assume that 
\begin{equation}
n,K,\NGibbs,\Ngrid\gg1\quad\mbox{with}\quad n\gg K.
\label{eq:largeDimnConds}
\end{equation}
Under (\ref{eq:largeDimnConds}) the singular value decomposition
of an $n\times K$ matrix requires $O(nK^2)$ operations and the 
singular value decomposition
of a $K\times K$ matrix requires $O(K^3)$ operations. This
leads to the following total operations order of magnitude statements
for each of Algorithms \ref{alg:dirGibbsPenSpl} and 
\ref{alg:odmGibbsPenSpl}:

\begin{center}
\begin{tabular}{lcc}
                          & Gibbs loop operations    & other operations     \\[1ex]
Algorithm \ref{alg:dirGibbsPenSpl}: & $O(\NGibbs K^3)$ & $O(nK^2) + O(\Nkept\Ngrid\,K)$   \\[1ex]
Algorithm \ref{alg:odmGibbsPenSpl}: & $O(\NGibbs K)$ & $O(nK^2)+O(\Ngrid K^2)+O(\Nkept\Ngrid\,K)$ 
\end{tabular}
\end{center}

For the Gibbs loops operations it is clear that 
Algorithm \ref{alg:odmGibbsPenSpl} provides
a two orders of magnitude improvement as a function
of $K$. This comes at the price of the extra $O(nK^2)$
operations required to decompose $\bX$ and $\bZ$
prior to the Algorithm \ref{alg:odmGibbsPenSpl}
Gibbs loop and the $O(\Ngrid K^2)$ step to compute
$\bZbreve_{\bg}$.
For typical values of $\NGibbs$ and $n$
the Gibbs loop speed-ups will outweigh the
cost of these non-loop steps. The constants
associated with the orders of magnitude also
impact the relative speeds of the two 
algorithms. Next, in Section \ref{sec:NPRcompExper}, 
we investigate the actual speed-ups via computer
experiments.

\subsection{Computer Experiments}\label{sec:NPRcompExper}

We coded Algorithms \ref{alg:dirGibbsPenSpl} and \ref{alg:odmGibbsPenSpl}
in the \texttt{C++} computer programming language and generated
100 replications of (\ref{eq:GaussNPReg}) for each of $n\in\{100,200,400\}$.
The number of basis functions was varied over $K\in\{25,50\}$
and the Gibbs sample sizes were fixed at $\Nburn=\Nkept=1000$.
The number of grid points was $\Ngrid=101$.
All calculations were performed on the first author's
\textsf{MacBook Air} computer which has 24 gigabytes of random
access memory and a 3.5 gigahertz processor.

\begin{figure}[t]
\centering
{\includegraphics[width=\textwidth]{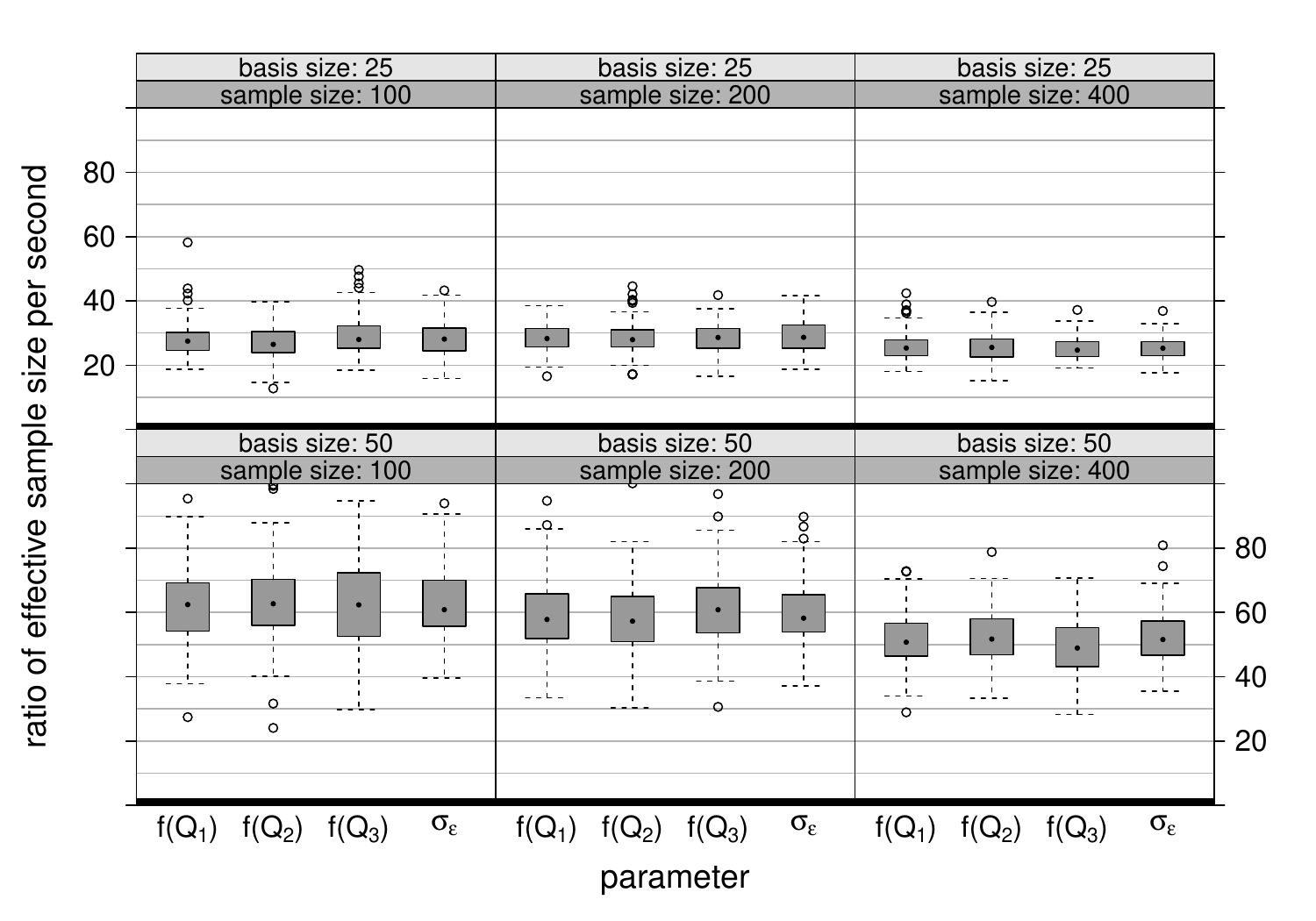}}
\caption{\textit{Side-by-side boxplots of the effective
sample size per second ratios for the computer
experiment involving Bayesian nonparametric regression
described in the text. The ratio numerator corresponds 
to Algorithm \ref{alg:odmGibbsPenSpl}.
The quantities of interest
are $f(Q_k)$, $k=1,2,3$, where  $Q_k$ is
the $k$th population quantile of the predictor
distribution and $\sigeps$ is the error standard
deviation. Each ratio corresponds to the
effective sample size per second for the direct
approach (Algorithm \ref{alg:odmGibbsPenSpl}) divided
by the same quantity for the orthogonalized 
design matrices approach (Algorithm \ref{alg:dirGibbsPenSpl}).
The thick horizontal lines correspond to a ratio of $1$.
}}
\label{fig:ESSpersNPR}
\end{figure}

Figure \ref{fig:ESSpersNPR} displays side-by-side boxplots 
of the effective sample size per second ratios for the kept Gibbs sample 
of four quantities of interest: $f$ evaluated at each of the population quantiles,
$f(Q_k)$, $k=1,2,3$, and the error standard deviation $\sigeps$.
The ratio numerator corresponds to Algorithm \ref{alg:odmGibbsPenSpl}.
Effective sample sizes of Markov chain Monte Carlo
samples are based on established approaches that account for
loss of information due to autocorrelation.
The particular version used here corresponds to the
\texttt{monitor()} function within the 
\textsf{R} package \textsf{rstan} (Stan Development Team, 2025)
with details given in that package's reference manual.

%

The side-by-side boxplots in Figure \ref{fig:ESSpersNPR}
reveal the considerable practical benefits of use
of orthogonalized design matrices. For the vast
majority of replications there are at least 
20-fold improvements in effective sample size per
second when the number of basis functions is $K=25$.
When the number of basis functions doubles to $K=50$
the improvement also approximately doubles and 
Algorithm \ref{alg:odmGibbsPenSpl} is usually
30--60 times faster than Algorithm \ref{alg:dirGibbsPenSpl}
in delivering the same quality of Gibbs samples.
For larger $n$ the improvement lessens slightly, 
which can be attributed to the 
$O(nK^2)$ singular value decomposition of $\bZ$ 
prior in Algorithm \ref{alg:odmGibbsPenSpl} 
prior to the Gibbs sampling phase. Nevertheless,
the improvement factors are still around 50.
A more formal exemplification of the improvements
is provided by Table \ref{tab:ESSpersNPR}, which
lists the medians and 95\% confidence intervals
for the basis size: 50 and sample size: 200 panel
of Figure \ref{fig:ESSpersNPR}.

\begin{table}[h]
\begin{center}
\begin{tabular}{lcccc}
                    & $f(Q_1)$ & $f(Q_2)$ & $f(Q_3)$ & $\sigeps$ \\[0.1ex]
\hline\\[-0.9ex]
CI low. & 56.2 & 55.6 &  58.7 & 57.6 \\[0.25ex]
median  & 57.8 & 57.3 &  60.8 & 58.2 \\[0.25ex]
CI upp. & 60.7 & 60.2 &  63.2 & 61.4 \\[0.25ex]
\hline
\end{tabular}
\end{center}
\caption{\textit{Medians and 95\% Wilcoxon confidence interval (CI) lower and
upper limits based on ratio of effective sample sizes per second data 
corresponding to the basis size: 50 and sample size: 200 panel of 
Figure \ref{fig:ESSpersNPR}.}}
\label{tab:ESSpersNPR}
\end{table}

Even though Table \ref{tab:ESSpersNPR} gives formal 
evidence of a big statistically significant improvement 
due to use of orthogonalized design matrices, we contend that 
that side-by-side boxplot summaries better convey the improvements. 
The remaining computer studies will only use such summaries.

\section{Generalized Additive Models}\label{sec:GAM}

We now consider the case of multiple predictors and
generalized additive models extensions of 
Bayesian nonparametric regression. Throughout
this section the data are assumed to be of the
form
\begin{equation}
(\bxLini,\bxNoni,y_i),\quad 1\le i\le n,
\label{eq:GAMnotation}
\end{equation}
where $y_i$ is the $i$th response observation.
For each $i$, $\bxLini$ is a vector of
observations corresponding to the $\dLin\times1$ 
vector of predictors $\bxLin$. The predictors in 
$\bxLin$ are assumed to have linear impacts 
on the mean response which, for example, is appropriate
for components of $\bxLin$ that are indicator variables.
The other predictors, corresponding to the $\dNon\times1$ 
vector of predictors $\bxNon$, are such that its
variables may have nonlinear impacts on the mean
response. For each $i$, $\bxNoni$ is a vector of
observations corresponding to $\bxNon$.

To illustrate the notation defined in the previous paragraph, 
consider the Boston mortgages applications data described in 
Section 1.3.2 of Harezlak \textit{et al.} (2018).
These data consist of several variables for 
$n=2,380$ mortgage applications in Boston, U.S.A.
An example probit additive model for these data is
\begin{eqnarray*}
&&\texttt{deny}_i\simind 
\mbox{Bernoulli}\Big(
\Phi\big(\beta_0+\beta_1\,\texttt{self-employed}_i
+\beta_2\,\texttt{single}_i+\beta_3\,\texttt{condominium}_i\\[-0.90ex]
&&\qquad\qquad\qquad\qquad\qquad\quad
+f_1(\texttt{DIR}_i)+f_2(\texttt{LVR}_i)
\big)
\Big),\quad 1\le i\le 2,380
\end{eqnarray*}
where $\texttt{deny}_i$ equals 1 if the $i$th mortgage application
was denied and 0 otherwise. The predictor observations
$\texttt{self-employed}_i$, $\texttt{single}_i$ and 
$\texttt{condominium}_i$ correspond to similarly defined indicator variables
for whether the applicant is self-employed, the applicant is single
and the property is a condominium, respectively.
Lastly, $\texttt{DIR}_i$ and $\texttt{LVR}_i$ are, respectively,
the debt to income ratio and the loan to property 
value ratio for the $i$th application. For this example, 
$\dLin=3$, $\dNon=2$ and the vectors in (\ref{eq:GAMnotation}) are
$$\bxLini=\left[\begin{array}{c}
\texttt{self-employed}_i\\
\texttt{single}_i\\
\texttt{condominium}_i
\end{array}  
\right]
\quad\mbox{and}\quad
\bxNoni=\left[\begin{array}{c}
\texttt{DIR}_i\\
\texttt{LVR}_i
\end{array}  
\right].
$$
In addition, $y_i=\texttt{deny}_i$.

The generic forms of the generalized additive models considered
in this section are
$$y_i\sim\left\{\begin{array}{ll} 
N(\eta_i,\sigeps^2), & y_i\ \mbox{continuous},\\[0.1ex]
\mbox{Bernoulli}\big(\Phi(\eta_i)\big), & y_i\ \mbox{binary},
\end{array}
\right.
$$
where
$$
\eta_i\equiv\beta_0+\sum_{j=1}^{\dLin}\beta_j\xLinji+\sum_{j=1}^{\dNon}f_j(\xNonji)
\quad\mbox{with}\quad
f_j(\xNonji)\equiv\beta_{\dLin+j}\,\xNonji+\sum_{k=1}^{K_j} u_{jk} z_{jk}(\xNonji)
$$
being a penalized spline model for the $j$th predictor in $\bxNon$. The $u_{jk}$
and $z_{jk}$ notation is analogous to that used in Section \ref{sec:bayNonParReg},
with the addition of the $j$ subscript to denote the $j$th predictor entering
the model non-linearly.

The design matrices for this generalized additive model set-up are
$$\bX\equiv\left[1\ \bxLiniT\ \bxNoniT\right]_{1\le i\le n}\quad
\mbox{and}\quad \bZ_j\equiv\big[\relstack{z_{jk}
(\xNonji)}{1\le k\le K}\big]_{1\le i\le n},\quad 1\le j\le\dNon.
$$
A Bayesian Gaussian response generalized additive model 
is
\begin{equation}
\begin{array}{c}
\by|\bbeta,\bu_1,\ldots,\bu_{\dNon}, 
\sigsqeps \sim N\Big(\bX\bbeta+{\displaystyle\sum_{j=1}^{\dNon}}
\bZ_j\,\bu_j,\sigeps^2\,\bI_n\Big),
\quad
\bbeta\sim N\big(\bzero,\sigmabeta^2\bI_{1+\dLin+\dNon}\big),\\[1ex]
\bu_j|\sigma_{uj}^2
\simind N(\bzero,\sigma_{uj}^2\bI_{K_j}),
\ \sigma_{uj}^{-2}|b_{uj}\simind\mbox{Gamma}\big(\smhalf,b_{uj}\big),
\ b_{uj}\simind\mbox{Gamma}(\smhalf,\su^{-2}),\\[1ex]
\quad 1\le j\le\dNon,
\quad\sigeps^{-2}|\beps\sim\mbox{Gamma}(\smhalf,\beps),
\quad\beps\sim\mbox{Gamma}(\smhalf,\seps^{-2}).
\end{array}
\label{eq:BayesGaussGAM}
\end{equation}
The binary response case involves replacement of the
first distributional statement in (\ref{eq:BayesGaussGAM})
by 
\begin{equation}
\begin{array}{c}
\balpha\ (n\times 1)\ \mbox{has $i$th entry $\alpha_i$
such that $y_i=I\big(\alpha_i\ge0\big)$,\quad $1\le i\le n$,}\\[0ex]
\balpha|\bbeta,\bu_1,\ldots,\bu_{\dNon}
\sim N\Big(\bX\bbeta+{\displaystyle\sum_{j=1}^{\dNon}}
\bZ_j\,\bu_j,\bI_n\Big)
\end{array}
\label{eq:BayesBernGAM}
\end{equation}
and removal of the $\sigeps^2$ and $\beps$ variables.
An appropriate orthogonalized design matrix reparametrization
of the linear predictor vector is
$$\bX\bbeta+\sum_{j=1}^{\dNon}\bZ_j\,\bu_j
=\bXbreve\bbetabreve+\sum_{j=1}^{\dNon}\bZbreve_j\,\bubreve_j
$$
where $\bXbreve$ and $\bbetabreve$ are defined as in 
Section \ref{sec:bayNonParReg}. For each $1\le j\le\dNon$,
$$\bZbreve_j\equiv\bUZj\diag(\bdZj)\quad\mbox{and}\quad
\bubreve_j\equiv \bVZj^T\bu_j
\quad\mbox{where}\quad
\bZ_j=\bUZj\diag(\bdZj)\bVZj^T
$$
is the singular value decomposition of $\bZ_j$
with $\bUZj$ being $n\times K_j$, $\bdZj$ being $K_j\times 1$
and $\bVZj$ being $K_j\times K_j$ such that
$\bUZj^T\bUZj=\bVZj^T\bVZj=\bVZj\bVZj^T=\bI_{K_j}$.

We are now ready to list Algorithm \ref{alg:odmGAMGibbs},
which describes Gibbs sampling for fitting the above Bayesian
generalized additive model with orthogonalized
design matrices speed-ups. It uses the following 
notation:
\begin{equation}
\begin{tabular}{l}
$K_j\equiv\mbox{the number of columns in}\ \bZbreve_j,\ 1\le j\le \dNon$,\\[1ex]
$\gothicc\ \mbox{is the}\ (\dNon+1)\times1  \mbox{vector with entries} 
\ \gothicc_1\equiv0 \mbox{and}\ \gothicc_{j+1}\equiv\sum_{k=1}^j K_k,
\ 1\le j\le \dNon$,\\[1ex]
$\ZbreveTyadj^{\langle j\rangle}
\equiv\mbox{the sub-block of}\ \ZbreveTyadj\  
\mbox{corresponding to rows}\ (\gothicc_j+1)\  
\mbox{to}\ \gothicc_{j+1},\ 1\le j\le \dNon$,\\[1ex]
$\ZbreveTXbrevej
\equiv\mbox{the sub-block of}\ \ZbreveTXbreve\ \mbox{corresponding to rows}
\ (\gothicc_j+1)\ \mbox{to}\ \gothicc_{j+1},\ 1\le j\le\dNon$,\\[1ex]
$\ZbreveTZbrevejjd
\equiv\mbox{the sub-block of}\ \ZbreveTZbreve
\ \mbox{corresponding to rows}\ (\gothicc_j+1)\ \mbox{to}\ \gothicc_{j+1}$,\\[1ex]
$\qquad\qquad\quad\ \mbox{and columns}\ (\gothicc_{j'}+1)\  \mbox{to}\ 
\gothicc_{j'+1},\ 1\le j,j'\le\dNon.$\\[1ex]
$\bX_{\bg}\ \mbox{and}\ \bZ_{j,\bg}, 1\le j\le\dNon,
\ \mbox{are grid-wise versions of the design for plotting grids of}$\\[1ex]
$\qquad\qquad\quad\ \mbox{size}\ \Ngrid$,\\[1ex]
$\bbetaLin\equiv\mbox{the}\ \dLin\times1\ \mbox{vector corresponding 
to the coefficients of the}\ \bxLini$.
\end{tabular}
\label{eq:PaulBugden}
\end{equation}

\begin{algorithm}[!t]
\begin{center}
\begin{minipage}[t]{155mm}
\hrule
\begin{small}
\begin{itemize}
\setlength\itemsep{0pt}
\item[] Inputs: $\by$ $(n\times 1)$,\ $\bX$ $\big(n\times(1+\dLin+\dNon)\big)$,
\ $\bZ_1$ $(n\times K_1)$,\ldots,$\bZ_{\dNon}$ $(n\times K_{\dNon})$,\\
\null$\qquad\qquad\bX_{\bg}$ $\big(\Ngrid\times(1+\dLin+\dNon)\big)$,
\ $\bZ_{1,\bg}$ $(\Ngrid\times K_1)$,\ldots,
$\bZ_{\dNon,\bg}$ $(\Ngrid\times K_{\dNon})$,\\
\null$\qquad\qquad\sigmabeta,\su,\seps>0$,\ $\Nburn,\Nkept\in\naturalNumbers$,\ 
$\textbf{responseType}\in\{\mbox{Gaussian},\mbox{Bernoulli}\}.$
\item[] Decompose $\bX=\bUX\diag\big(\bdX\big)\bVX^T$ where 
$\bUX^T\bUX=\bVX^T\bVX=\bVX\bVX^T=\bI_{1+\dLin+\dNon}$
\item[] $\bXbreve\longleftarrow\bUX\diag(\bdX)$\ \ ;\ \ 
$\bdXsq\longleftarrow\bdX\odot\bdX$\ \ ;\ \ 
$\XbreveTyadj\longleftarrow\bXbreveT\by$

\item[] For $j=1,\ldots,\dNon$:
\begin{itemize}
\setlength\itemsep{0pt}
\item[] Decompose $\bZ_j=\bUZj\diag\big(\bdZj\big)\bVZj^T$ where 
$\bUZj^T\bUZj=\bVZj^T\bVZj=\bVZj\bVZj^T=\bI_{K_j}$
\item[] $\bZbreve_j\longleftarrow\bUZj\diag(\bdZj)$
\ \ ;\ \ $\bdZjsq\longleftarrow\bdZj\odot\bdZj$
\end{itemize}
\item[] $\bZbreve\longleftarrow[\bZbreve_1\cdots\bZbreve_{\dNon}]$
\ \ ;\ \ $\ZbreveTyadj\longleftarrow\bZbreveT\by$
\ \ ;\ \ $\ZbreveTXbreve\longleftarrow\bZbreveT\bXbreve$
\ \ ;\ \ $\ZbreveTZbreve\longleftarrow\bZbreveT\bZbreve$
\item[] Initialize: $(\sigeps^{-2})^{[0]},\beps>0$\ \ ;\ \ 
For each $1\le j\le\dNon$:
$\bubreve^{[j]}$ $(K_j\times 1)$, $(\sigma_{uj}^{-2})^{[0]},b_{uj}>0$ 
\item[] For $s=1,\ldots,\Nburn+\Nkept$:
\begin{itemize}
\setlength\itemsep{0pt}
\item[] $\brbetabreve\longleftarrow 
\XbreveTyadj-{\displaystyle\sum_{j=1}^{\dNon}}
\ZbreveTXbrevejT\,\bubreve_j^{[s-1]}$\ \ ;\ \ 
$\bnubetabreve\longleftarrow(\sigeps^{-2})^{[s-1]}
\bdX^2+\sigmabeta^{-2}\bone_{1+\dLin+\dNon}$
\item[] $\bzbetabreve\sim N(\bzero,\bI_{1+\dLin+\dNon})$\ \ \ ;\ \ \ 
$\bbetabreve^{[s]}\longleftarrow {\displaystyle\frac{\bzbetabreve}
{\sqrt{\bnubetabreveFORsqrt}}}+{\displaystyle\frac{(\sigeps^{-2})^{[s-1]}
\brbetabreve}{\bnubetabreve}}$
\item[] For $j=1,\ldots,\dNon$: $\bubrevejCurr\longleftarrow\bubreve_j^{[s-1]}$
\item[] For $j=1,\ldots,\dNon$: 
\begin{itemize}
\setlength\itemsep{0pt}
\item[] $\brujbreve\longleftarrow 
\ZbreveTyadj^{\langle j\rangle}
-\ZbreveTXbrevej\,\bbetabreve^{[s]}
-{\displaystyle\sum_{j'\ne j}^{\dNon}}
\ZbreveTZbrevejjd\bubrevejdCurr$
\item[] $\bnuujbreve\longleftarrow(\sigeps^{-2})^{[s-1]}
\bdZj^2+(\sigma_{uj}^{-2})^{[s-1]}\bone_{K_j}$
\item[] $\bzujbreve\sim N(\bzero,\bI_{K_j})$\ \ \ ;\ \ \ 
$\bubrevejCurr\longleftarrow {\displaystyle\frac{\bzujbreve}
{\sqrt{\bnuujbreve}}}+{\displaystyle\frac{
(\sigeps^{-2})^{[s-1]}
\brujbreve}{\bnuujbreve}}$
\end{itemize}
\setlength\itemsep{0pt}
\item[] For $j=1,\ldots,\dNon$: $\bubreve_j^{[s]}\longleftarrow\bubrevejCurr$
\item[] For $j=1,\ldots,\dNon$: 
\begin{itemize}
\setlength\itemsep{0pt}
\item[] $(\sigma_{uj}^{-2})^{[s]}\longleftarrow\GammaDist\Big(\smhalf(K_j+1),b_{uj}+
\smhalf\big\Vert\bubreve_j^{[s]}\big\Vert^2\Big)$
\ \ ;\ \ $b_{uj}\longleftarrow\GammaDist\Big(1,(\sigma_{uj}^{-2})^{[s]}+
\su^{-2}\Big)$
\end{itemize}
\item[] $\bdeta\longleftarrow\bXbreve\bbetabreve^{[s]}
+{\displaystyle\sum_{j=1}^{\dNon}}\bZbreve_j\bubreve_j^{[s]}$
\item[] If $\textbf{responseType}\ \mbox{is Gaussian}$ then
\begin{itemize}
\setlength\itemsep{0pt}
\item[] $(\sigeps^{-2})^{[s]}\longleftarrow\GammaDist\Big(\smhalf(n+1),\beps+
\smhalf\Vert\by-\bdeta\Vert^2)$\ ; 
$\beps\longleftarrow\GammaDist\Big(1,(\sigeps^{-2})^{[s]}+\seps^{-2}\Big)$
\end{itemize}
\item[] If $\textbf{responseType}\ \mbox{is Bernoulli}$ then
\begin{itemize}
\setlength\itemsep{0pt}
\item[] $(\sigeps^{-2})^{[s]}\longleftarrow 1$\ ;
\ $\zeta\sim\mbox{Truncated-Normal}_+\big((2y_i-1)
\eta_i,1\big)$
\ \ ;\ \ $\alpha_i\longleftarrow (2y_i-1)\zeta$
\item[] $\XbreveTyadj\longleftarrow\bXbreveT\balpha$\ \ ;\ \ 
$\ZbreveTyadj\longleftarrow\bZbreveT\balpha$
\end{itemize}
\end{itemize}
\item[] \textit{continued on a subsequent page $\ldots$}
\end{itemize}
%
%
\end{small}
\hrule
\end{minipage}
\end{center}
\caption{\textit{An orthogonalized design matrices Gibbs
sampling scheme for the Bayesian generalized additive models 
(\ref{eq:BayesGaussGAM}) and (\ref{eq:BayesBernGAM}).}}
\label{alg:odmGAMGibbs}
\end{algorithm}
%

\setcounter{algorithm}{3}
\begin{algorithm}[!t]
\begin{center}
\begin{minipage}[t]{155mm}
\hrule
\begin{small}
\begin{itemize}
\setlength\itemsep{0pt}
\item[] $\bXbreve_{\bg}\longleftarrow\bX_{\bg}\bVX$ 
\ \ \ ;\ \ \ For $j=1,\ldots,\dNon: \bZbreve_{j,\bg}\longleftarrow \bZ_{j,\bg}\bVZj$
\item[] For $s=1,\ldots,\Nkept$:
\begin{itemize}
\setlength\itemsep{0pt}
\item[] $\bbetaBreveTEMP\longleftarrow 
\mbox{entries $2$ to $(1+\dLin)$ of}\ \bbetabreve^{[s+\Nburn]}$
\ \ ;\ \ 
$\bbetaLins\longleftarrow\bVX\bbetaBreveTEMP$ 
\item[] For $j=1,\ldots,\dNon$:
\begin{itemize}
\item[] $\bxgcurr\longleftarrow\mbox{$(1+\dLin+j)$th column of}\  
\bXbreve_{\bg}$\ \ ;\ \ 
$\betabrevecurr\longleftarrow\mbox{$(1+\dLin+j)$th entry of}\ 
\bbetabreve^{[s+\Nburn]}$
\item[] $\bfhat_{j\bg}^{[s]}\longleftarrow
\bxgcurr\betabrevecurr
+\bZbreve_{j,\bg}\bubreve_j^{[s+\Nburn]}$
\end{itemize}
\end{itemize}
\item[] Outputs: $\big\{\bbetaLins,\bfhat_{j\bg}^{[s]},
(\sigma_{uj}^{-2})^{[s]}:1\le s\le \Nkept,\ 1\le j\le\dNon\big\}$
\item[] If $\textbf{responseType}\ \mbox{is Gaussian}$ then also output
$\big\{(\sigeps^{-2})^{[s]}:1\le s\le \Nkept\big\}$
\end{itemize}
\end{small}
\hrule
\end{minipage}
\end{center}
\caption{\textbf{continued.}\ \textit{This is a continuation of the description of 
this algorithm that commences on a preceding page.}}
\end{algorithm}
%

\begin{figure}[t]
\centering
{\includegraphics[width=\textwidth]{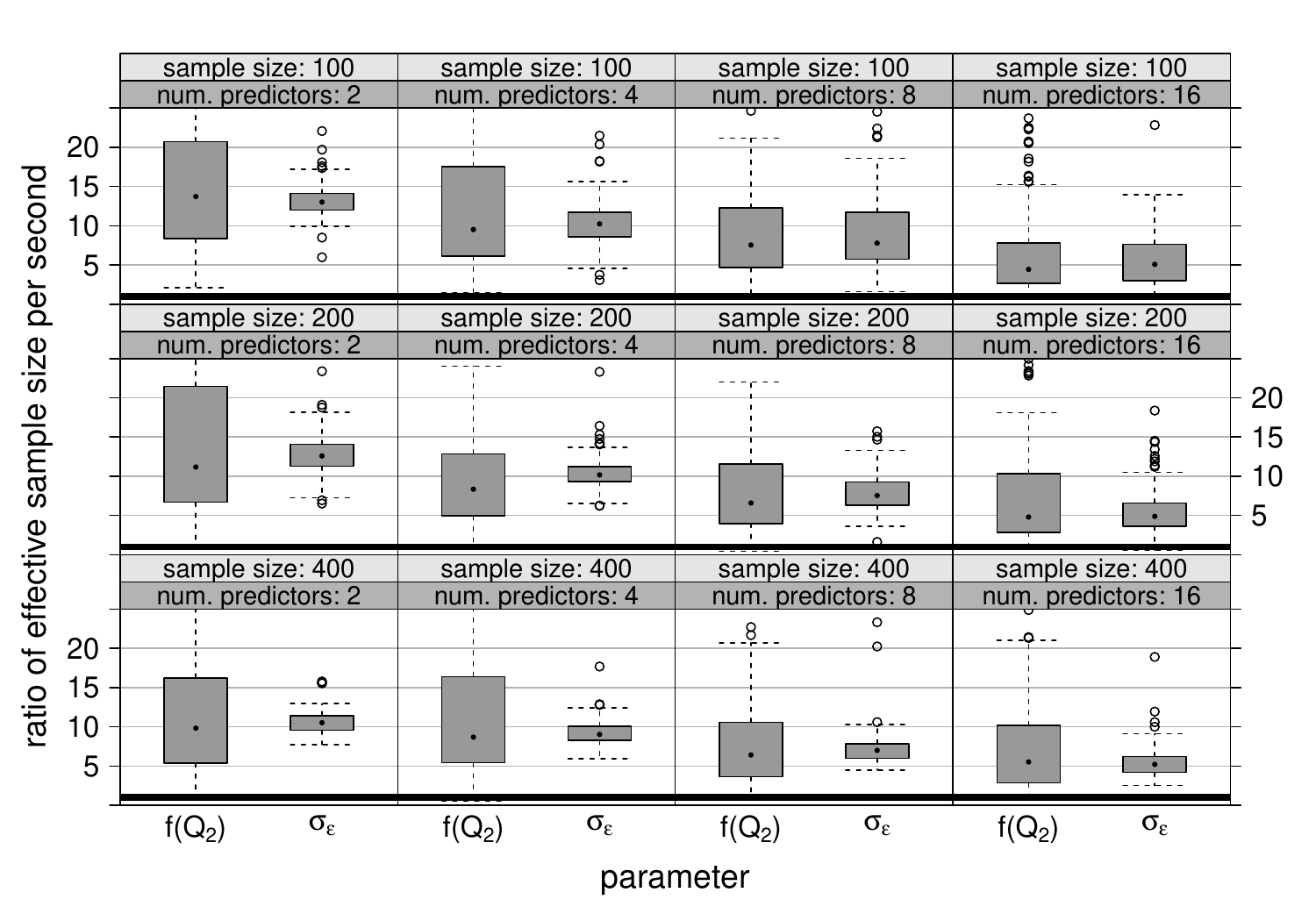}}
\caption{\textit{Side-by-side boxplots of the effective
sample size per second ratios for the computer
experiment involving Bayesian generalized additive models
described in the text. Each ratio corresponds to the
effective sample size per second for the orthogonalized 
design matrices approach divided by the same quantity for the direct approach. 
The thick horizontal lines correspond to a ratio of $1$.}}
\label{fig:ESSpersGAM}
\end{figure}

We ran a computer experiment to compare the practical performance
of Algorithm \ref{alg:odmGAMGibbs} with the direct computation
alternative in the case of Gaussian responses. The sample sizes 
ranged over $n\in\{100,200,400\}$ and the number of predictors 
entering the model non-linearly ranged over $\dNon\in\{2,4,8,16\}$. 
The  number of predictors entering the model linearly was fixed at $\dLin=0$
and number of basis functions for each predictor was fixed at $K=25$. 
The chains that were monitored are the vertical slice
corresponding to the population median of all $\dNon$ predictors,
which we denote by $f(Q_2)$, and the error standard deviation
$\sigeps$. Figure \ref{fig:ESSpersGAM} summarises the effective
sample size per second ratios using side-by-side boxplots.
It shows that effective sample sizes per second are
about $5$--$20$ times larger when the orthogonalized
design matrices approach is used. The advantage tends
to decrease for larger models but there are still
around $5$-fold improvements for generalized additive
models with $16$ predictors.

%

\section{Group-Specific Curves Models}\label{sec:GSCmodels}

The group-specific curves models (e.g.\ Donnelly \textit{et al.} 1995)
that we consider here are based on grouped data of the form
$$(x_{ij},y_{ij}),\quad 1\le j\le n_i,\ 1\le i\le m,$$
where, for example, $x_{ij}$ is the $j$th predictor measurement
within the $i$th group. The number of groups is $m$.
The models have generic forms
\begin{equation}
y_{ij}\sim\left\{\begin{array}{ll} 
N\big(f(x_{ij})+g_i(x_{ij}),\sigeps^2\big), & y_{ij}\ \mbox{continuous},\\[0.1ex]
\mbox{Bernoulli}\big(\Phi\big(f(x_{ij})+g_i(x_{ij})\big)\big), & y_{ij}\ \mbox{binary},
\end{array}
\right.
\label{eq:genericGSC}
\end{equation}
for smooth functions $f$ and $g_i$, $1\le i\le m$.
The function $f$ models the global mean response, whereas $g_i$ 
models the deviation from the global mean response for group $i$.

As in this article's previous sections, mixed model-based penalized
splines can be used to formulate a hierarchical Bayesian model
for (\ref{eq:genericGSC}).
For $1\le i\le m$ define
$$\bx_i\equiv[x_{ij}]_{1\le j\le n_i},\quad
\by_i\equiv[y_{ij}]_{1\le j\le n_i},\quad
\bX_i\equiv[\bone_{n_i}\ \bx_i],
$$
$$
\bZgbli\equiv\big[
\zgblo(\bx_i)\cdots\zgblKgbl(\bx_i)\big]
\quad\mbox{and}\quad
\bZgrpi\equiv\big[
\zgrpo(\bx_i)\cdots\zgrpKgrp(\bx_i)
\big]
$$
where $\{\zgblk(\cdot):1\le k\le\Kgbl\}$ 
and $\{\zgrpk(\cdot):1\le k\le\Kgrp\}$ 
are spline basis functions for the global $f$ and 
group-specific $g_i$ functions, respectively. 
The full Gaussian response Bayesian group-specific
curves model that we consider here is 

\begin{equation}
{\setlength\arraycolsep{1pt}
\begin{array}{c}
\by_i|\bbeta,\buGbl,\buLini,\buGrpi,\sigeps^2\simind N\big(\bX_i\bbeta
+\bZgbli\buGbl+\bX_i\buLini
+\bZgrpi\buGrpi,\sigeps^2\bI_{n_i}\big),\ 1\le i\le m,\\[1ex]
\buGbl|\sigmaGbl^2\sim N(\bzero,\sigmaGbl^2\bI_{\Kgbl}),
\quad \buLini|\bSigma\simind N(\bzero,\bSigma),\quad 
\buGrpi|\sigmaGrp^2\simind N(\bzero,\sigmaGrp^2\bI_{\Kgrp}),\ 1\le i\le m,\\[1ex]
\bbeta\sim N(\bzero,\sigmabeta^2\bI_2),\quad\sigeps^2|\beps\sim\mbox{Gamma}(\smhalf,\beps),
\quad\beps\sim\mbox{Gamma}(\smhalf,\seps^{-2}),\\[1ex]
\quad\sigmaGbl^{-2}|\bGbl\sim\mbox{Gamma}(\smhalf,\bGbl),
\quad\bGbl\sim\mbox{Gamma}(\smhalf,\sGbl^{-2}),\\[1ex]
\quad\sigmaGrp^{-2}|\bGrp\sim\mbox{Gamma}(\smhalf,\bGrp),
\quad\bGrp\sim\mbox{Gamma}(\smhalf,\sGrp^{-2}),\\[1ex]
\bSigma^{-1}|\bLino,\bLint\sim\mbox{Wishart}\left(3,4
\left[\begin{array}{cc}    
\bLino   & 0      \\
  0      & \bLint
\end{array}
\right]\right),\quad 
\bLino,\bLint\simind\mbox{Gamma}(\smhalf,\sLin^{-2})
\end{array}
}
\label{eq:UncleToby}
\end{equation}
for hyperparameters $\sigmabeta,\seps,\sGbl,\sGrp,\sLin>0$.
The notation $\bX\sim\mbox{Wishart}(\kappa,\bLambda)$ signifies
that $\bX$ has a Wishart distribution with shape parameter 
$\kappa$ and rate matrix $\bLambda$ and is such that the 
corresponding density function satisfies 
$\pDens(\bX)\propto |\bX|^{(\kappa-3)/2}\exp\{-\smhalf\tr(\bLambda\bX)\}$
for $\bX$ symmetric and positive definite.
The last line of (\ref{eq:UncleToby}) corresponds to $\bSigma$
having a marginally non-informative prior distribution, for sufficiently
large $\sLin$, as described in Huang \myand Wand (2013).

For the second case of (\ref{eq:genericGSC}), for which the
$y_{ij}$s are binary, the Albert \myand Chib (1993) 
approach involves the introduction of the auxiliary variables,
$\alpha_{ij}$, $1\le i\le m$, $1\le j\le n_i$, such that 
\begin{equation}
y_{ij}\equiv I(\alpha_{ij}>0)\quad\mbox{and the vectors}\quad
\balpha_i\equiv[\alpha_{ij}]_{1\le j\le n_i},\quad 1\le i\le m.
\label{eq:turningGreen}
\end{equation}
With (\ref{eq:turningGreen}) in place, the full binary response 
Bayesian group-specific curves model is the same as (\ref{eq:UncleToby}) 
but with the first line replaced by 
$$\balpha_i|\bbeta,\buGbl,\buLini,\buGrpi\simind N\big(\bX_i\bbeta
+\bZgbli\buGbl+\bX_i\buLini
+\bZgrpi\buGrpi,\bI_{n_i}\big),\quad 1\le i\le m,
$$

We now present Algorithm \ref{alg:odmGSCGibbs} for Gibbs sampling with 
orthogonalized design matrices speed-ups, and let the matrices $\bXbreve_i$ 
and $\bZgbliBreve$, $1\le i\le m$, be defined according to 
$$\bXbreve=\left[\begin{array}{c}
\bXbreve_1\\
\vdots\\
\bXbreve_m
\end{array}
\right]
 \quad\mbox{and}\quad 
\bZgblBreve=
\left[\begin{array}{c}
\bZgbloneBreve\\
\vdots\\
\bZgblmBreve
\end{array}
\right]$$
with the number of rows in $\bXbreve_i$ and $\bZgbliBreve$ being 
the same as the number of rows in $\by_i$. The grid-wise design matrices
$\bX_{\bg}$, $\bZgblg$ and  $\bZgrpg$ are also inputted
to Algorithm \ref{alg:odmGSCGibbs}.

\begin{algorithm}[!t]
\begin{center}
\begin{minipage}[t]{155mm}
\hrule
\begin{small}
\begin{itemize}
\setlength\itemsep{0pt}
\item[]
\item[] Inputs: $\ \ \by_i$ $(n_i\times 1)$,\ $\bX_i$ $(n_i\times2)$,
\ $\bZgbli$ $(n_i\times \Kgbl)$,\ $\bZgrpi$ $(n_i\times \Kgrp)$,\ $1\le i\le m$,\\
\null$\qquad\qquad \bX_{\bg}$ $(\Ngrid\times2)$,
\ $\bZgblg$ $(\Ngrid\times \Kgbl)$,\ $\bZgrpg$ $(\Ngrid\times \Kgrp)$,\\
\null$\qquad\qquad\sigmabeta,\seps,\sGbl,\sLin,\sGrp>0$,\ $\Nburn,\Nkept\in\naturalNumbers$,\ 
$\textbf{responseType}\in\{\mbox{Gaussian},\mbox{Bernoulli}\}$.
\item[] Decompose $\bX=\bUX\diag\big(\bdX\big)\bVX^T$ where 
$\bUX^T\bUX=\bVX^T\bVX=\bVX\bVX^T=\bI_2.$
\item[] Decompose $\bZgbl=\bUZgbl\diag\big(\bdZgbl\big)\bVZgbl^T$ where\\
\null$\qquad\qquad\quad\ \ \bUZgbl^T\bUZgbl=\bVZgbl^T\bVZgbl =\bVZgbl\bVZgbl^T =\bI_{\Kgbl}.$
\item[] $\bXbreve\longleftarrow\bUX\diag(\bdX)$
\ \ ;\ \ $\bZgblBreve\longleftarrow\bUZgbl\diag\big(\bdZgbl\big)$
\ \ ;\ \ $\bdXsq\longleftarrow\bdX\odot\bdX$
\ \ ;\ \ $\bdZgblsq\longleftarrow$$\bdZgbl$$\odot$$\bdZgbl$
\item[] $\XTyiadj\longleftarrow\bX_i^T\by_i$
\ \ ;\ \ $\XTXi\longleftarrow\bX_i^T\bX_i$
\ \ ;\ \ $\XTXbrevei\longleftarrow\bX_i^T\bXbreve_i$
\ \ ;\ \ $\XTZgbliBreve\longleftarrow\bX_i^T\bZgbliBreve$
\item[] $\SUMXbreveTyadj\longleftarrow \displaystyle\sum_{i=1}^{m}\bXbreve_i^T\by_i$
\ ;\ $\SUMZgblBreveTyadj\longleftarrow \displaystyle\sum_{i=1}^{m}\bZgbliBreve^T\by_i$
\ ;\ $\SUMXbreveTZgblBreve\longleftarrow \displaystyle\sum_{i=1}^{m}\bXbreve_i^T\bZgbliBreve$
\item[] For $i=1,\ldots,m$:
\begin{itemize}
\setlength\itemsep{0pt}
\item[] Decompose $\bZgrpi=\bUZgrpi\diag\big(\bdZgrpi\big)\bVZgrpi^T$ 
where $\bUZgrpi^T\bUZgrpi=\bI_{\Kgrp}$\\
\null and $\bVZgrpi^T\bVZgrpi=\bVZgrpi\bVZgrpi^T=\bI_{\Kgrp}$
\item[] $\bZgrpiBreve\longleftarrow\bUZgrpi\diag\big(\bdZgrpi\big)$
\ \ ;\ \ $\bdZgrpisq\longleftarrow$ $\bdZgrpi$ $\odot$ $\bdZgrpi$
\end{itemize}
\item[] $\ZgrpBreveTyiadj\longleftarrow\bZgrpiBreve^T\by_i$
\ ;\ $\XbreveTZgrpiBreve\longleftarrow\bXbreve_i^T\bZgrpiBreve$
\ ;\ $\ZgblBreveTZgrpiBreve\longleftarrow\bZgbliBreve^T\bZgrpiBreve$
\item[] Initialize: $\ \ \bugblBreve^{[0]}\ (\Kgbl\times 1)$, 
\ $\buLini^{[0]}\ (2\times m)$,
\ $\buGrpiBreve^{[0]}\ (\Kgrp\times 1)$,\\
\null$\qquad\qquad\quad
(\sigeps^{-2})^{[0]},(\sigmaGbl^{-2})^{[0]},
(\bSigma^{-1})_{11}^{[0]},(\bSigma^{-1})_{22}^{[0]},(\sigmaGrp^{-2})^{[0]},
\beps,\bGbl,\bLino,\bLint,\bGrp>0$,\\
\null$\qquad\qquad\quad \XTyiadj\ (2\times 1)$
,\ $\SUMXbreveTyadj\ (2\times 1)$
,\ $\SUMZgblBreveTyadj\ (\Kgbl\times 1)$
,\ $\ZgrpBreveTyiadj\ (\Kgrp\times 1)$.
\item[] For $s=1,\ldots,\Nburn+\Nkept$:
\begin{itemize}
\setlength\itemsep{0pt}
\item[] $\brbetabreve\longleftarrow 
\SUMXbreveTyadj 
- \SUMXbreveTZgblBreve\,\bugblBreve^{[s-1]} 
- {\displaystyle\sum_{i=1}^{m}}(\XTXbrevei)^T\buLini^{[s-1]} 
- {\displaystyle\sum_{i=1}^{m}}\XbreveTZgrpiBreve\buGrpiBreve^{[s-1]}$
\item[] $\bnubetabreve\longleftarrow(\sigeps^{-2})^{[s-1]}
\bdX^2+\sigmabeta^{-2}\bone_2$
\item[] $\bzbetabreve\sim N(\bzero,\bI_2)$\ \ \ ;\ \ \ 
$\bbetabreve^{[s]}\longleftarrow {\displaystyle\frac{\bzbetabreve}
{\sqrt{\bnubetabreveFORsqrt}}}+{\displaystyle\frac{(\sigeps^{-2})^{[s-1]}
\brbetabreve}{\bnubetabreve}}$
\item[] $\brugblbreve\longleftarrow 
\SUMZgblBreveTyadj 
- (\SUMXbreveTZgblBreve)^T\,\bbetabreve^{[s]} 
- {\displaystyle\sum_{i=1}^{m}}(\XTZgbliBreve)^T\buLini^{[s-1]} 
- {\displaystyle\sum_{i=1}^{m}}\ZgblBreveTZgrpiBreve\buGrpiBreve^{[s-1]}$
\item[] $\bnuugblbreve\longleftarrow(\sigeps^{-2})^{[s-1]}
\bdZgblsq+(\sigmaGbl^{-2})^{[s-1]}\bone_{\Kgbl}$
\item[] $\bzugblbreve\sim N(\bzero,\bI_{\Kgbl})$
\ \ \ ;\ \ \ 
$\bugblBreve^{[s]}\longleftarrow {\displaystyle\frac{\bzugblbreve}
{\sqrt{\bnuugblbreve}}}+{\displaystyle\frac{(\sigeps^{-2})^{[s-1]}
\brugblbreve}{\bnuugblbreve}}$
\item[] For $i=1,\ldots,m$: 
\begin{itemize}
\setlength\itemsep{0pt}
\item[] $\bruLini\longleftarrow \XTyiadj
- \XTXbrevei\,\bbetabreve^{[s]}
- \XTZgbliBreve\bugblBreve^{[s]}
- \XTZgrpiBreve\buGrpiBreve^{[s]}$
\item[] $\MATRIXuLini\longleftarrow (\sigeps^{-2})^{[s-1]}\XTXi+(\bSigma^{-1})^{[s-1]}$
\item[] Decompose $\MATRIXuLini=\bUuLini\diag(\bduLini)\bUuLini^T$ where 
$\bUuLini^T\bUuLini=\bUuLini\bUuLini^T=\bI_2$ 
\item[] $\bzuLini\sim N(\bzero,\bI_2)$
\ \ \ ;\ \ \ $\buLini^{[s]}\longleftarrow \bUuLini
\left({\displaystyle \frac{\bUuLini^T\bzuLini}{\sqrt{\bduLini}}} 
+ {\displaystyle \frac{(\sigeps^{-2})^{[s-1]}\bUuLini^T\bruLini}{\bduLini}}\right)$
\end{itemize}
\end{itemize}
\item[] \textit{continued on a subsequent page $\ldots$}
\end{itemize}
\end{small}
\hrule
\end{minipage}
\end{center}
\caption{\textit{An orthogonalized design matrices Gibbs
sampling scheme for the group-specific curves models 
(\ref{eq:UncleToby}) and (\ref{eq:turningGreen}).}}
\label{alg:odmGSCGibbs}
\end{algorithm}
%

\setcounter{algorithm}{4}
\begin{algorithm}[!t]
\begin{center}
\begin{minipage}[t]{155mm}
\hrule
\begin{small}
\begin{itemize}
\setlength\itemsep{0pt}
\item[]
\item[]
\begin{itemize}
\setlength\itemsep{0pt}
\item[] For $i=1,\ldots,m$: 
\begin{itemize}
\setlength\itemsep{0pt}
\item[] $\brugrpibreve\longleftarrow 
\ZgrpBreveTyiadj 
- (\XbreveTZgrpiBreve)^T\,\bbetabreve^{[s]} 
- (\ZgblBreveTZgrpiBreve)^T\bugblBreve^{[s]}
- (\XTZgrpiBreve)^T\buLini^{[s]} 
$
\item[] $\bnuugrpibreve\longleftarrow(\sigeps^{-2})^{[s-1]}
\bdZgrpisq+(\sigmaGrp^{-2})^{[s-1]}\bone_{\Kgrp}$
\item[] $\bzugrpibreve\sim N(\bzero,\bI_{\Kgrp})$
\ \ \ ;\ \ \ $\buGrpiBreve^{[s]}\longleftarrow 
{\displaystyle \frac{\bzugrpibreve}{\sqrt{\bnuugrpibreve}}} 
+{\displaystyle \frac{(\sigeps^{-2})^{[s-1]}\brugrpibreve}{\bnuugrpibreve}}$
\end{itemize}
\item[] $\bdeta_i\longleftarrow\bXbreve_i\bbetabreve^{[s]}
+\bZgbliBreve\bugblBreve^{[s]}
+\bX_i\buLini^{[s]}
+\bZgrpiBreve\buGrpiBreve^{[s]}$,
\ $i=1,\ldots,m$
\item[] If $\textbf{responseType}\ \mbox{is Gaussian}$ then
\begin{itemize}
\setlength\itemsep{0pt}
\item[] $(\sigeps^{-2})^{[s]}\longleftarrow\GammaDist
\Big(\smhalf\Big({\displaystyle\sum_{i=1}^m}n_i+1\Big), 
\beps+\smhalf{\displaystyle\sum_{i=1}^m}\Vert\by_i-\bdeta_i\Vert^2\Big)$
\item[] $\beps\longleftarrow\GammaDist\Big(1,(\sigeps^{-2})^{[s]}+\seps^{-2}\Big)$
\end{itemize}
\item[] If $\textbf{responseType}\ \mbox{is Bernoulli}$ then
\begin{itemize}
\setlength\itemsep{0pt}
\item[] $(\sigeps^{-2})^{[s]}\longleftarrow 1$
\item[] For $1\le i\le m$:
\begin{itemize}
\item[] For $1\le j\le n_i$:\\
\null$\qquad \zeta\sim\mbox{Truncated-Normal}_+ \big((2y_{ij}-1)(\bdeta_i)_j,1\big)$
\ ;\ $\alpha_{ij}\longleftarrow (2y_{ij}-1)\zeta$
\end{itemize}
\item[] $\XTyiadj\longleftarrow\bX_i^T\balpha_i$
\ \ ;\ \ $\SUMXbreveTyadj\longleftarrow \displaystyle\sum_{i=1}^{m}\bXbreve_i^T\balpha_i$
\item[] $\SUMZgblBreveTyadj\longleftarrow \displaystyle\sum_{i=1}^{m}\bZgbliBreve^T\balpha_i$
\ ;\ $\ZgrpBreveTyiadj\longleftarrow\bZgrpiBreve^T\balpha_i$
\end{itemize}
\item[] $(\sigmaGbl^{-2})^{[s]}\longleftarrow\GammaDist\Big(\smhalf(\Kgbl+1),\bGbl+
\smhalf\big\Vert\bugblBreve^{[s]}\big\Vert^2\Big)$
\ \ ;\ \ $\bGbl\longleftarrow\GammaDist\Big(1,(\sigmaGbl^{-2})^{[s]}+
\sGbl^{-2}\Big)$
\item[] $(\bSigma^{-1})^{[s]}\longleftarrow\mbox{Wishart}\Big(m+3, 
4\diag(\bLino,\bLint) + \displaystyle\sum_{i=1}^{m}\buLini^{[s]}\buLini^{[s]T}\Big)$
\item[] $\bLino\longleftarrow\GammaDist\Big(2,2(\bSigma^{-1})_{11}^{[s]}+
\sLin^{-2}\Big)$
\ \ ;\ \ $\bLint\longleftarrow\GammaDist\Big(2,2(\bSigma^{-1})_{22}^{[s]}+
\sLin^{-2}\Big)$
\item[] $(\sigmaGrp^{-2})^{[s]}\longleftarrow\GammaDist\Big(\smhalf(m\Kgrp+1), 
\bGrp+\smhalf\displaystyle\sum_{i=1}^{m} \big\Vert\buGrpiBreve^{[s]}\big\Vert^2\Big)$
\item[]$\bGrp\longleftarrow\GammaDist\Big(1,(\sigmaGrp^{-2})^{[s]}+
\sGrp^{-2}\Big)$
\end{itemize}
\item[] $\bXbreve_{\bg}\longleftarrow \bX_{\bg}\bVX$
\ \ \ ;\ \ \ $\bZbrevegblg\longleftarrow \bZgblg\bVZ$
\ \ \ ;\ \ \ For $i=1,\ldots,m$: 
$\bZbrevegrpig\,{\leftarrow\!\!\!-\!\!\!-}\,\bZgrpg\bVZgrpi$
\item[] For $s=1,\ldots,\Nkept$:
\begin{itemize}
\setlength\itemsep{0pt}
\item[] $\bdetahat_{\bg}^{[s]}\longleftarrow\bXbreve_{\bg}\bbetabreve^{[s+\Nburn]}
+\bZbrevegblg\bugblBreve^{[s+\Nburn]}$
\item[] For $i=1,\ldots,m$:
\begin{itemize}
\item[] $\bdetahat_{i\bg}^{[s]}\longleftarrow \bdetahat_{\bg}^{[s]}
+\bX_{\bg}\buLini^{[s+\Nburn]}+\bZbrevegrpig\buGrpiBreve^{[s+\Nburn]}$
\end{itemize}
\item[] $(\sigmaGbl^{-2})^{[s]}\longleftarrow (\sigmaGbl^{-2})^{[s+\Nburn]}$\ \ ;\ \ 
$(\bSigma^{-1})^{[s]}\longleftarrow (\bSigma^{-1})^{[s+\Nburn]}$\ \ ;\ \ 
$(\sigmaGrp^{-2})^{[s]}\longleftarrow (\sigmaGrp^{-2})^{[s+\Nburn]}$
\item[] If \textbf{responseType} is Gaussian then
$(\sigeps^{-2})^{[s]}\longleftarrow (\sigeps^{-2})^{[s+\Nburn]}$
\end{itemize}
\item[] Outputs: $\big\{\bdetahat_{\bg}^{[s]},
\ \bdetahat_{i\bg}^{[s]},\ (\sigmaGbl^{-2})^{[s]},\ (\bSigma^{-1})^{[s]}
,\ (\sigmaGrp^{-2})^{[s]}:1\le s\le \Nkept,\ 1\le i\le m\big\}$.
\item[] If \textbf{responseType} is Gaussian then also output
$\big\{(\sigeps^{-2})^{[s]}:1\le s\le \Nkept\big\}$.
\end{itemize}
\end{small}
\hrule
\end{minipage}
\end{center}
\caption{\textbf{continued.}\ \textit{This is a continuation of the description of 
this algorithm that commences on a preceding page.}}
\end{algorithm}
%

\begin{figure}[t]
\centering
{\includegraphics[width=\textwidth]{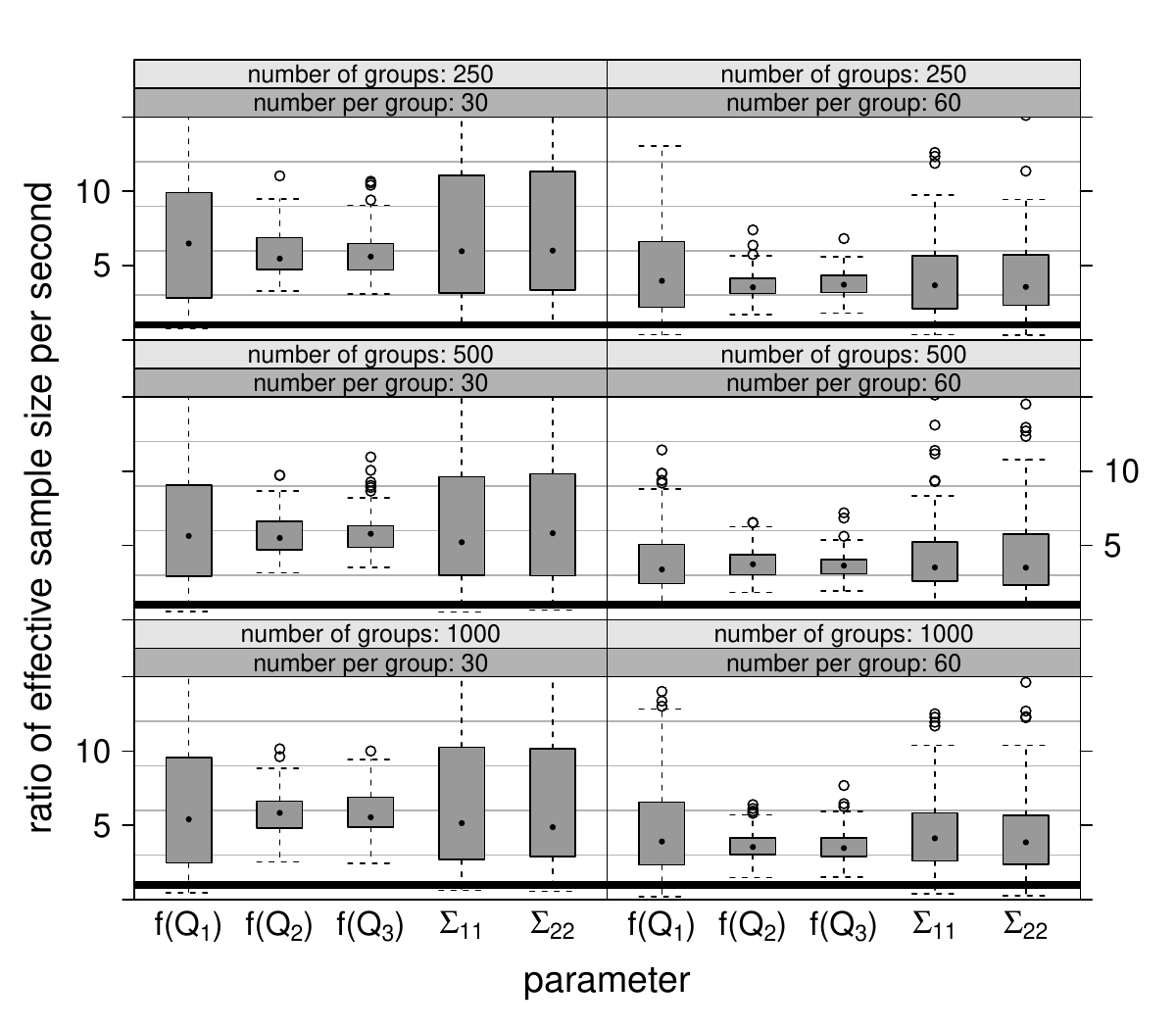}}
\caption{\textit{
Side-by-side boxplots of the effective
sample size per second ratios for the computer
experiment involving Bayesian group-specific curves 
models described in the text. Each ratio corresponds to the
effective sample size per second for the orthogonalized 
design matrices approach divided by the same quantity for the direct approach. 
The thick horizontal lines correspond to a ratio of $1$.}}
\label{fig:ESSpersGSC}
\end{figure}

Figure \ref{fig:ESSpersGSC} shows the results of a computer experiment
in which synthetic data were generated according to the binary
response version of (\ref{eq:genericGSC}) with the number of 
groups ranging over $m\in\{250,500,1000\}$ and the number 
of observations within the $i$th group fixed to be $n_i=n$
with $n\in\{30,60\}$. The spline basis sizes were fixed at
$\Kgbl=25$ and $\Kgrp=15$ and the Gibbs sample sizes
were $\Nburn=\Nkept=5,000$.
The quantities of interest are $f$ evaluated at each 
of the population quantiles, $f(Q_k)$, $k=1,2,3$, and the diagonal
entries of $\bSigma$, $\Sigma_{kk}$, $k=1,2$.
The side-by-side boxplots in Figure \ref{fig:ESSpersGSC} 
show that the orthogonalized design matrices approach improves
upon the direct approach by factors of around 5--15.

\subsection{Application to Adolescent Somatic Growth Data}

\begin{figure}[t]
\centering
{\includegraphics[width=0.9\textwidth]{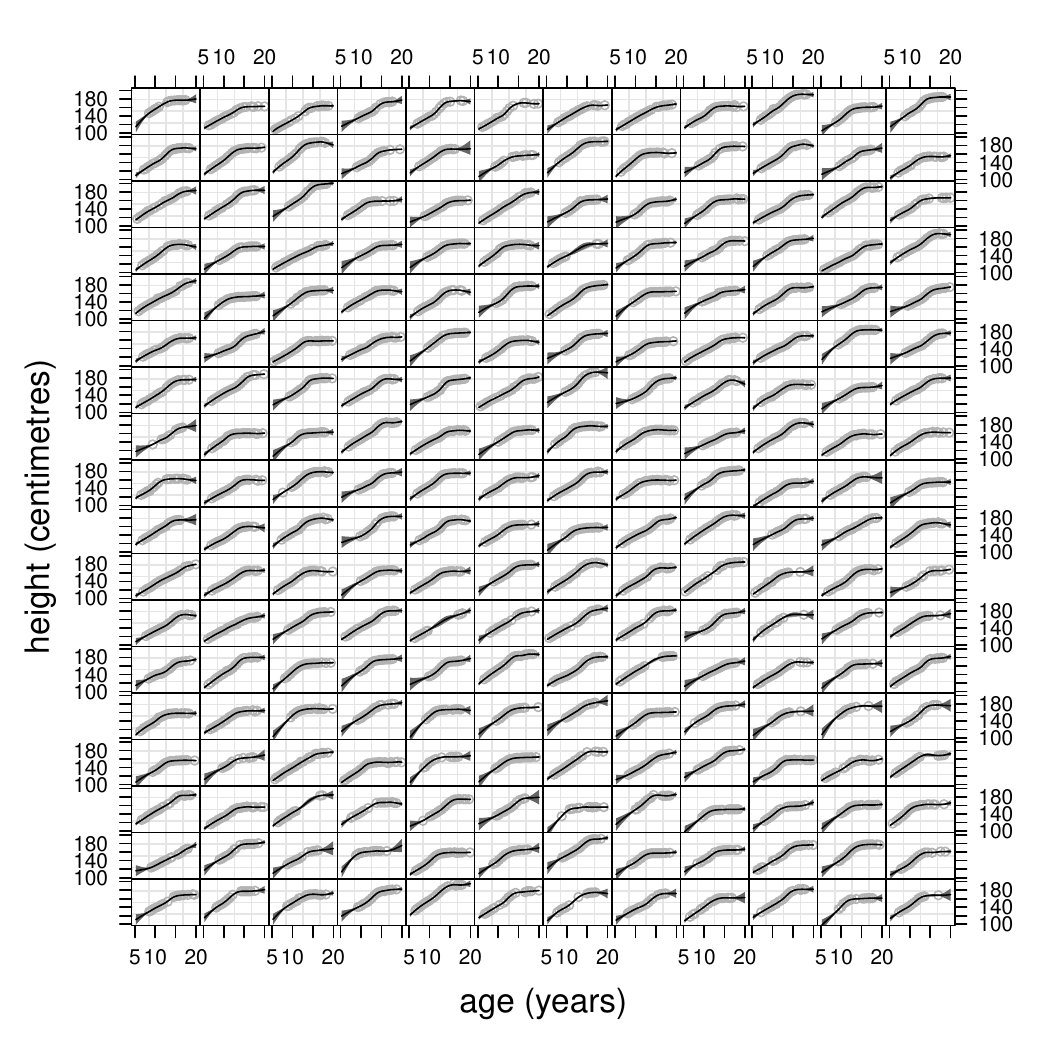}}
\caption{\textit{Somatic growth data for each of 
216 adolescents from Indiana, U.S.A., 
from  the study described in Pratt \textit{et al.} (1989).
The curves are posterior means and the shading indicates
pointwise 95\% credible intervals 
for the Bayesian group-specific curves model described in
Section \ref{sec:GSCmodels} and obtained using Algorithm \ref{alg:odmGSCGibbs}.}}
\label{fig:growthIndianaFits}
\end{figure}

We applied the Gaussian response version
of Algorithm \ref{alg:odmGSCGibbs} and its direct
counterpart to data on adolescent somatic growth from 
the study described in Pratt \textit{et al.} (1989). The data 
are part of the \textsf{R} package \textsf{HRW} 
(Harezlak \textit{et al.} 2021) and stored in the 
data frame \texttt{growthIndiana}. It consists of 9 or more
longitudinal height measurements taken approximately every six months
for each of 216 adolescents from Indiana, U.S.A., 

Figure \ref{fig:growthIndianaFits} shows the Bayesian group-specific
curve model fits. The basis sizes are $\Kgbl=25$ and $\Kgrp=9$ and the 
hyperparameters values were set to
$\sigmabeta=\seps=\sGbl=\sLin=\sGrp=10^5$
after global standardisation of the age
and height data before input into Algorithm \ref{alg:odmGSCGibbs}.
The Gibbs sample size values are $\Nburn=\Nkept=10,000$.
In each panel of Figure \ref{fig:growthIndianaFits} 
the curve corresponds to the posterior mean of 
$f(\texttt{age})+g_i(\texttt{age})$, $1\le i\le 216$,
(in the notation of (\ref{eq:genericGSC}))
and the shaded region corresponds to pointwise 95\%
credible intervals after back-transformation
to the original units. Simple enhancements of model
(\ref{eq:UncleToby}) and Algorithm \ref{alg:odmGSCGibbs}
could be used to make inferences concerning ethnicity
and gender contrasts. Of interest here is the 
speed-ups afforded by use of orthogonalized design
matrices.

When run on the first author's \textsf{MacBook Air} computer,
with specifications given in Section \ref{sec:NPRcompExper},
the direct approach took 51.8 seconds
whereas Algorithm \ref{alg:odmGSCGibbs} only
took 10.8 seconds. This  approximately $5$-fold
speed-up exemplifies the advantages of orthogonalized
design matrices in applications.

\section{Variational Inference}\label{sec:varInf}

Orthogonalized design matrices speed-ups also apply
to other iterative Bayesian fitting and inference
procedures. Here we provide a flavor of how the
principle used in the previous few sections for Gibbs sampling 
also applies to variational inference.

Consider again the Bayesian nonparametric regression model
(\ref{eq:BayePenSplineModel}). Suppose that
we approximate the joint posterior density 
function of the model parameters:
\begin{equation}
\pDens(\bbeta,\bu,b_u,\beps,\sigma_u^2,\sigeps^2|\by)
\quad\mbox{by the product density form}\quad 
\qDens(\bbeta,\bu,b_u,\beps)\qDens(\sigma_u^2,\sigeps^2).
\label{eq:prodRestric}
\end{equation}
Using results laid out in, for example, Chapter 10 of Bishop (2006)
the optimal $\qDens$-densities in terms of minimizing
the Kullback-Leibler divergence of the first
density function in (\ref{eq:prodRestric}) from the second are 
$$\qDens^*(\bbeta,\bu,b_u,\beps)=\qDens^*(\bbeta,\bu)\qDens^*(b_u)
\qDens^*(\beps)\quad\mbox{and}\qquad
\qDens^*(\sigma_u^2)\qDens^*(\sigeps^2)
$$
where the $\qDens^*$-densities have forms
$$
\begin{array}{l}
\qDens^*(\bbeta,\bu): N(\bmu_{\qDens(\bbeta,\bu)},\Sigma_{\qDens(\bbeta,\bu)}),
\ 
\qDens^*(b_u): \GammaDist(\kappa_{\qDens(b_u)},\lambda_{\qDens(b_u)}),\ 
\qDens^*(\beps): \GammaDist(\kappa_{\qDens(\beps)},\lambda_{\qDens(\beps)}),\\[0.5ex]
\qDens^*(\sigma^2_u):\ 
\mbox{Inverse-Gamma}(\kappa_{\qDens(\sigma_u^2)},\lambda_{\qDens(\sigma_u^2)})
\quad\mbox{and}\quad\qDens^*(\sigeps^2): 
\mbox{Inverse-Gamma}(\kappa_{\qDens(\sigeps^2)},\lambda_{\qDens(\sigeps^2)}).
\end{array}
$$
The optimal parameters of the $\qDens^*$-densities can be 
obtained using the coordinate ascent iterative algorithm
listed in Algorithm \ref{alg:dirMFVBpenSpl}.
The stopping criterion depends on the approximate
marginal log-likelihood under product restriction 
(\ref{eq:prodRestric}), which we denote by 
$\log\{\punder(\by;\qDens)\}$.
Its explicit form is given later in this section.
The approximate posterior density function, 
of the grid-wise fit values vector 
$\bfhat_{\bg}\equiv \bX_{\bg}\bbeta+\bZ_{\bg}\bu$
is Multivariate Normal with mean vector
$E_{\qDens}\big(\bfhat_{\bg}\big)=\bC_{\bg}\bmu_{\qDens(\bbeta,\bu)}$
and covariance matrix 
$\Cov_{\qDens}\big(\bfhat_{\bg}\big)=\bC_{\bg}\bSigma_{\qDens(\bbeta,\bu)}\bC_{\bg}^T$.
Each of these matrices is outputted by Algorithm \ref{alg:dirMFVBpenSpl} to
facilitate approximate Bayesian inference for $f$.

\setcounter{algorithm}{5}
\begin{algorithm}[!th]
\begin{center}
\begin{minipage}[t]{155mm}
\hrule
\begin{small}
\begin{itemize}
\item[] Inputs: $\by$ $(n\times 1)$,\ $\bX$ $(n\times2)$,\ $\bZ$ $(n\times K)$,\
\ $\bX_{\bg}$ $(\Ngrid\times2)$,\ $\bZ_{\bg}$ $(\Ngrid\times K)$
$\sigmabeta,\su,\seps,\epsToler>0$ 
\item[] Initialize: $\mu_{\qDens(\sigma_u^{-2})},\mu_{\qDens(b_u)},
\mu_{\qDens(\sigeps^{-2})},\mu_{\qDens(\beps)}>0$\ \ ;\ \ 
$\kappa_{\qDens(\sigma_u^2)}\longleftarrow\smhalf(K+1)$,\ \ ;\ \ 
$\kappa_{\qDens(\sigeps^2)}\longleftarrow\smhalf(n+1)$
\item[] $\bC\longleftarrow[\bX\ \bZ]$\ \ ;\ \ 
$\bC_{\bg}\longleftarrow [\bX_{\bg}\ \bZ_{\bg}]$\ \ ;\ \ 
$\CTy\longleftarrow \bC^T\by$\ \ ;\ \ $\CTC\longleftarrow \bC^T\bC$
\item[] Cycle:
\begin{itemize}
\item[] 
$\bOmegaqbetau\longleftarrow\mu_{\qDens(\sigeps^{-2})}\CTC
+\diag\big(\big[\sigmabeta^{-2}\bone_2^T\ 
\mu_{\qDens(\sigma_u^{-2})}\bone_K^T\big]^T\big)$
\item[] Decompose 
$\bOmegaqbetau=\bUqbetau\diag(\bdqbetau)\bUqbetau^T$ where 
\\
$\bUqbetau^T\bUqbetau=\bUqbetau\bUqbetau^T=\bI_{K+2}$
\item[] $\bSigmaqbetau\longleftarrow 
\bUqbetau\diag(\bone/\bdqbetau)\bUqbetau^T$
\ \ ;\ \ 
$\bmuqbetau\longleftarrow\mu_{\qDens(\sigeps^{-2})}
\bSigmaqbetau\CTy$
\item[] $\bmuqu\longleftarrow\mbox{lower $K\times1$ block of}\ \bmuqbetau$\ \ ;\ \ 
$\bSigmaqu\longleftarrow\mbox{lower right $K\times K$ block of}\ \bSigmaqbetau$\
\item[] $\lambda_{\qDens(\sigma_u^2)}\longleftarrow 
\mu_{\qDens(b_u)}+\smhalf\Vert\bmuqu\Vert^2
+\smhalf\tr\big(\bSigmaqu\big)$\ \ ;\ \ 
$\mu_{\qDens(\sigma_u^{-2})}\longleftarrow
\kappa_{\qDens(\sigma_u^2)}\big/\lambda_{\qDens(\sigma_u^2)}$
\item[] $\lambda_{\qDens(b_u)}\longleftarrow 
\mu_{\qDens(\sigma_u^{-2})}+\su^{-2}$\ \ ;\ \ 
$\mu_{\qDens(b_u)}\longleftarrow 1\big/\lambda_{\qDens(b_u)}$
\item[] $\lambda_{\qDens(\sigeps^2)}\longleftarrow 
\mu_{\qDens(\beps)}+\smhalf\big\Vert\by-\bC\bmuqbetau\big\Vert^2
+\smhalf\tr\big(\CTC\,\bSigmaqbetau\big)$
\ \ ;\ \ 
$\mu_{\qDens(\sigeps^{-2})}\longleftarrow
\kappa_{\qDens(\sigeps^2)}\big/\lambda_{\qDens(\sigeps^2)}$
\item[] $\lambda_{\qDens(\beps)}\longleftarrow 
\mu_{\qDens(\sigeps^{-2})}+\seps^{-2}$\ \ ;\ \ 
$\mu_{\qDens(\beps)}\longleftarrow 1\big/\lambda_{\qDens(\beps)}$
\end{itemize}
\item[] until the relative change in $\log\{\punder(\by;\qDens)\}$
is less than $\epsToler$.
\item[] $E_{\qDens}\big(\bfhat_{\bg}\big)\longleftarrow \bC_{\bg}\bmu_{\qDens(\bbeta,\bu)}$
\ \ ;\ \ $\Cov_{\qDens}\big(\bfhat_{\bg}\big)\longleftarrow 
\bC_{\bg}\bSigma_{\qDens(\bbeta,\bu)}\bC_{\bg}^T$
\item[] Outputs: $\big\{E_{\qDens}\big(\bfhat_{\bg}\big),
\Cov_{\qDens}\big(\bfhat_{\bg}\big),
\kappa_{\qDens(\sigma_u^2)},\lambda_{\qDens(\sigma_u^2)}\
\kappa_{\qDens(\sigeps^2)},\lambda_{\qDens(\sigeps^2)}\big\}$
\end{itemize}
\end{small}
\hrule
\end{minipage}
\end{center}
\caption{\textit{A direct mean field variational Bayes algorithm for fitting
and inference for the Bayesian nonparametric regression model 
(\ref{eq:BayePenSplineModel}).}}
\label{alg:dirMFVBpenSpl}
\end{algorithm}
%

In Algorithm \ref{alg:dirMFVBpenSpl} note that an $O(nK^2)$ singular 
value decomposition, for inversion of $\bOmegaqbetau$, is carried out 
within each iteration. Algorithm \ref{alg:odmMFVBpenSpl} avoids
this computational cost by working with an orthogonalized version
of $\bC\equiv[\bX\ \bZ]$. The approximate marginal log-likelihood
with respect to the $(\bbetabreve,\bubreve)$ parameterization
is denoted by $\log\{{\widebreve\punder}(\by;\qDens)\}$.
An explicit expression is given later in this section.

\begin{algorithm}[!th]
\begin{center}
\begin{minipage}[t]{155mm}
\hrule
\begin{small}
\begin{itemize}
\item[] Inputs: $\by$ $(n\times 1)$,\ $\bX$ $(n\times2)$,\ $\bZ$ $(n\times K)$,
\ $\bX_{\bg}$ $(\Ngrid\times2)$,\ $\bZ_{\bg}$ $(\Ngrid\times K)$,
$\sigmabeta,\su,\seps,\epsToler>0$
\item[] Initialize: $\mu_{\qDens(\sigma_u^{-2})},\mu_{\qDens(b_u)},
\mu_{\qDens(\sigeps^{-2})},\mu_{\qDens(\beps)}>0$\ \ ;\ \ 
$\kappa_{\qDens(\sigma_u^2)}\longleftarrow\smhalf(K+1)$,\ \ ;\ \ 
$\kappa_{\qDens(\sigeps^2)}\longleftarrow\smhalf(n+1)$
\item[] $\bC\longleftarrow[\bX\ \bZ]$\ \ \ ;\ \ \ 
$\bC_{\bg}\longleftarrow [\bX_{\bg}\ \bZ_{\bg}]$
\item[] Decompose $\bC=\bUC\diag(\bdC)\bVC^T$ where $\bUC^T\bUC
=\bVC^T\bVC=\bVC\bVC^T=\bI_{K+2}$\\
$\bCbreve\longleftarrow\bUC\diag(\bdC)$\ \ ;\ \ 
$\CbreveTy\longleftarrow \bCbreveT\by$\ \ ;\ \ 
$\bdC^2\longleftarrow \bdC\odot\bdC$
\item[] Cycle: 
\begin{itemize}
\item[] 
$\bsigsqqbetaubreve
\longleftarrow \bone_{K+2}\Big/\Big(\mu_{\qDens(\sigeps^{-2})}\bdCsq
+\big[\sigmabeta^{-2}\bone_2^T\ 
\mu_{\qDens(\sigma_u^{-2})}\bone_K^T\big]^T\Big)$\ ;\ 
$\bmuqbetaubreve\longleftarrow\mu_{\qDens(\sigeps^{-2})}
\bsigsqqbetaubreve\CbreveTy$
\item[] $\bmuqubreve\longleftarrow\mbox{lower $K\times1$ block of}\ \bmuqbetaubreve$\ \ ;\ \ 
$\bsigsqqubreve\longleftarrow\mbox{lower $K\times 1$ block of}\ \bsigsqqbetaubreve$\
\item[] $\lambda_{\qDens(\sigma_u^2)}\longleftarrow 
\mu_{\qDens(b_u)}+\smhalf\Vert\bmuqubreve\Vert^2
+\smhalf\bone_K^T\bsigsqqubreve$\ \ ;\ \ 
$\mu_{\qDens(\sigma_u^{-2})}\longleftarrow
\kappa_{\qDens(\sigma_u^2)}\big/\lambda_{\qDens(\sigma_u^2)}$
\item[] $\lambda_{\qDens(b_u)}\longleftarrow 
\mu_{\qDens(\sigma_u^{-2})}+\su^{-2}$\ \ ;\ \ 
$\mu_{\qDens(b_u)}\longleftarrow 1\big/\lambda_{\qDens(b_u)}$
\item[] $\lambda_{\qDens(\sigeps^2)}\longleftarrow 
\mu_{\qDens(\beps)}+\smhalf\Big\Vert\by-\bCbreve\bmuqbetaubreve\Big\Vert^2
+\smhalf(\bdCsq)^T\bsigsqqbetaubreve$
\ \ ;\ \ 
$\mu_{\qDens(\sigeps^{-2})}\longleftarrow
\kappa_{\qDens(\sigeps^2)}\big/\lambda_{\qDens(\sigeps^2)}$
\item[] $\lambda_{\qDens(\beps)}\longleftarrow 
\mu_{\qDens(\sigeps^{-2})}+\seps^{-2}$\ \ ;\ \ 
$\mu_{\qDens(\beps)}\longleftarrow 1\big/\lambda_{\qDens(\beps)}$
\end{itemize}
\item[] until the relative change in $\log\{{\widebreve\punder}(\by;\qDens)\}$
is less than $\epsToler$.
\item[] $\bCbreve_{\bg}\longleftarrow \bC_{\bg}\bVC$\ \ ;\ \  
$E_{\qDens}\big(\bfhat_{\bg}\big)\longleftarrow \bCbreve_{\bg}\bmuqbetaubreve$
\ \ ;\ \ $\Cov_{\qDens}\big(\bfhat_{\bg}\big)\longleftarrow 
\bCbreve_{\bg}\diag\big(\bsigsqqbetaubreve\big)\bCbrevegT$
\item[] Outputs: $\big\{E_{\qDens}\big(\bfhat_{\bg}\big),
\Cov_{\qDens}\big(\bfhat_{\bg}\big),
\kappa_{\qDens(\sigma_u^2)},\lambda_{\qDens(\sigma_u^2)}\
\kappa_{\qDens(\sigeps^2)},\lambda_{\qDens(\sigeps^2)}\big\}$
\end{itemize}
\end{small}
\hrule
\end{minipage}
\end{center}
\caption{\textit{An orthogonalized design matrices speed-up of
Algorithm \ref{alg:dirMFVBpenSpl} for the Bayesian nonparametric
regression model (\ref{eq:BayePenSplineModel}).}}
\label{alg:odmMFVBpenSpl}
\end{algorithm}
%

\null\vfill\eject
For the stopping criteria in each of Algorithm \ref{alg:dirMFVBpenSpl}
and Algorithm \ref{alg:odmMFVBpenSpl} first define
{\setlength\arraycolsep{1pt}
\begin{eqnarray*}
\Tsc_{\qDens(\sigma_u^2,b_u,\sigeps^2,\beps)}
&\equiv&\mu_{\qDens(b_u)}\big(\mu_{\qDens(1/\sigma_u^2)}+s_u^{-2}\big)
+\smhalf(K+1)\log\big(\lambda_{\qDens(\sigma_u^2)}\big)
+\log\big(\lambda_{\qDens(b_u)}\big)
\\[0.1ex]
&&\quad +\mu_{\qDens(\beps)}\big(\mu_{\qDens(1/\sigeps^2)}+\seps^{-2}\big)
+\smhalf(n+1)\log\big(\lambda_{\qDens(\sigeps^2)}\big)
+\log\big(\lambda_{\qDens(\beps)}\big)
\end{eqnarray*}
}
and let `const' denote constant terms such as $-\smhalf\,n\log(2\pi)$.
Then expressions for the $\qDens$-density dependent
components of $\log\{\punder(\by;\qDens)\}$ and 
$\log\{{\widebreve\punder}(\by;\qDens)\}$ are:
{\setlength\arraycolsep{1pt}
\begin{eqnarray*}
\log\{\punder(\by;\qDens)\}&=&\smhalf\log|\bSigma_{\qDens(\bbeta,\bu)}|
-\smhalf\sigma_{\beta}^{-2}\big\{
\Vert\bmu_{\qDens(\bbeta)}\Vert^2+\tr(\bSigma_{\qDens(\bbeta)})\big\}\\[-0.4ex]
&&\quad
-\smhalf\mu_{\qDens(\sigma_u^{-2})}\big\{\Vert\bmu_{\qDens(\bu)}\Vert^2
+\tr(\bSigma_{\qDens(\bu)})\big\}
-\Tsc_{\qDens(\sigma_u^2,b_u,\sigeps^2,\beps)}+\mbox{const}
\end{eqnarray*}
}
and
{\setlength\arraycolsep{1pt}
\begin{eqnarray*}
\log\{{\widebreve\punder}(\by;\qDens)\}&=&\smhalf\bone_{K+2}^T
\log\big(\bsigsqqbetaubreve\big)
-\smhalf\sigma_{\beta}^{-2}\big(
\big\Vert\bmu_{\qDens(\bbetabreve)}\big\Vert^2+\bone_2^T\bsigsqqbetabreve
\big)\\[-0.3ex]
&&\quad
-\smhalf\mu_{\qDens(\sigma_u^{-2})}\big(\Vert\bmu_{\qDens(\bubreve)}\Vert^2+
\bone_K^T\bsigsqqubreve\big)-\Tsc_{\qDens(\sigma_u^2,b_u,\sigeps^2,\beps)}+\mbox{const}
\end{eqnarray*}
}
where, for example, $\bSigma_{\qDens(\bbeta)}$ is the upper left $2\times2$
block of $\bSigma_{\qDens(\bbeta,\bu)}$.

There are numerous extensions of Gaussian response
nonparametric regression for which the principles illustrated 
in this section apply. Some of these are the extensions described
in Sections \ref{sec:GAM} and \ref{sec:GSCmodels}. Others include
streamlined variational inference for higher level random
effects as described in Nolan \textit{et al.} (2020)
and multiply nested group-specific curves as described
in Menictas \textit{et al.} (2021).

\section{Conclusions}\label{sec:conclusion}

The orthogonalized design matrices approach to Bayesian semiparametric
regression is a small-cost adjustment that yields significant
speed-ups. Bayesian computing algorithms such as 
Gibbs sampling and coordinate ascent variational inference
require two orders of magnitude fewer operations. Our 
computer experiments show practical speed-ups as high
as factors exceeding $60$ and almost always by factors
of $5$ to $10$. In conclusion, of orthogonalized design matrices
has clear benefits in semiparametric regression
applications where speed is important.

\appendix

\section{Result 1 and Derivation}

Throughout this article we use a singular value decomposition
approach to obtain draws from Multivariate Normal distributions.
These are underpinned by Result 1, which we now state and prove.

\vskip2mm
\noindent
\textbf{Result 1}: \emph{Suppose that $\VECTOR\ (d\times1)$ and 
$\MATRIX\ (d\times d)$ are two matrices such $\MATRIX$ is 
symmetric and positive definite. Next, suppose}
$\MATRIX=\bU\diag(\bd)\bU^T$ \emph{is a decomposition such
that $\bU^T\bU=\bU\bU^T=\bI_d$. If}
$$\bz\sim N(\bzero,\bI_d)\quad\mbox{and} 
\quad\bx\equiv\bU\left(\frac{\bU^T\bz}{\sqrt{\bd}}+\frac{\bU^T\VECTOR}{\bd}\right)
\quad\mbox{then}
\quad\bx\sim N(\MATRIX^{-1}\VECTOR,\MATRIX^{-1}).
$$

\vskip2mm
\noindent
\textit{Proof of Result 1}.

\vskip2mm
First note that the mean of $\bx$ is
$$E(\bx)=\bU\left(\frac{\bU^T\VECTOR}{\bd}\right)
=\bU\diag(\bone/\bd)\bU^T\VECTOR
=\{\bU\diag(\bd)\bU^T\}^{-1}\VECTOR
=\MATRIX^{-1}\VECTOR.
$$
The covariance matrix of $\bx$ is
{\setlength\arraycolsep{1pt}
\begin{eqnarray*}
\Cov(\bx)=\Cov\Big(\bU\left(\frac{\bU^T\bz}{\sqrt{\bd}}\right)\Big)
&=&\Cov\Big(\bU\diag(\bone/\sqrt{\bd})\bU^T\bz\Big)\\[1ex]
&=&\bU\diag(\bone/\sqrt{\bd})\bU^T\Cov(\bz)
\bU\diag(\bone/\sqrt{\bd})\bU^T\\[1ex]
&=&\bU\diag(\bone/\bd)\bU^T
=\big\{\bU\diag(\bd)\bU^T\big\}^{-1}=\MATRIX^{-1}.
\end{eqnarray*}
}
Since $\bx$ is a linear transformation of $\bz$, it also has a 
Multivariate Normal distribution and Result 1 holds.

\section*{Acknowledgements}

This research was partially supported by the Australian
Research Council Discovery Project DP230101179,
the Indonesian Education Scholarship, Center for Higher Education Funding 
and Assessment and the Indonesian Endowment Fund for Education.

\section*{References}

\bib
Albert, J.H. \myand Chib, S. (1993). Bayesian analysis of binary and polychotomous
response data. \textit{Journal of the American Statistical Association},
\textbf{88}, 669--679.

\bib
Bishop, C.M. (2006). \textit{Pattern Recognition and Machine Learning.}
New York: Springer.

\bib
Demmler, A \myand Reinsch, C. (1975). Oscillation matrices with spline smoothing.
\textit{Numerische Mathematik}, \textbf{24}, 375--382.

\bib
Donnelly, C.A., Laird, N.M. and Ware, J.H. (1995).
Prediction and creation of smooth curves for
temporally correlated longitudinal data.
\textit{Journal of the American Statistical Association},
\textbf{90}, 984--989.

\bib
Gelman, A. (2006). Prior distributions for variance parameters 
in hierarchical models. \textit{Bayesian Analysis}, \textbf{1}, 515--533.

\bib
Harezlak, J., Ruppert, D. and Wand, M.P. (2018).
\textit{Semiparametric Regression with R}.
New York: Springer.  

\bib
Harezlak, J., Ruppert, D. and Wand, M.P. (2021).
\textsf{HRW 1.0}.
Datasets, functions and scripts for semiparametric regression 
supporting Harezlak, Ruppert \& Wand (2018).
\textsf{R} package. {\tt https://CRAN.R-project.org/package=HRW}

\bib
Huang, A. and Wand, M.P. (2013).
Simple marginally noninformative prior distributions 
for covariance matrices. \textit{Bayesian Analysis}, 
\textbf{8}, 439--452.

\bib
Menictas, M., Nolan, T.H., Simpson, D.G. and Wand, M.P. (2021).
Streamlined variational inference for higher level group-specific curve models.
\textit{Statistical Modelling}, \textbf{21}, 479--519.

\bib
Nolan, T.H., Menictas, M. and Wand, M.P. (2020).
Streamlined computing for variational inference with higher level 
random effects. \textit{Journal of Machine Learning Research}, 
\textbf{21(157)}, 1--62. 

\bib
Polson, N.G., Scott, J.G. \myand Windle, J. (2013).
Bayesian inference for logistic models using P\'olya-Gamma
latent variables. \textit{Journal of the American Statistical
Association}, \textbf{108}, 1339--1349.

\bib
Pratt, J.H., Jones, J.J., Miller, J.Z., Wagner, M.A. 
\myand Fineberg, N.S. (1989). 
Racial differences in aldosterone excretion and plasma aldosterone 
concentrations in children. \textit{New England Journal of Medicine}, 
\textbf{321}, 1152--1157.

\bib
Robert, C.P. (1995). Simulation of truncated normal variates.
\textit{Statistics and Computing}, \textbf{5}, 121--125.

\bib
Stan Development Team (2025). \textsf{rstan}: the \textsf{R}
interface to \textsf{Stan}. \textsf{R} package version
2.32.7. \texttt{https://mc-stan.org/}

\bib
Wand, M.P. and Ormerod, J.T. (2008).
On semiparametric regression with O'Sullivan penalized splines.
\textit{Australian and New Zealand Journal of Statistics},
\textbf{50}, 179--198.

\end{document}